\documentclass[review]{elsarticle}

\usepackage{lineno,hyperref}
\usepackage{color}
\usepackage{amsmath, amssymb} 
\usepackage{rotating}
\modulolinenumbers[5]

\journal{Journal of \LaTeX\ Templates}

\definecolor{cinnamon}{rgb}{0.82, 0.41, 0.12}

\begin{document}

\begin{frontmatter}

\title{
Channel flow of rigid sphere suspensions: \\ particle dynamics in the inertial regime
}



%
%

\author[address1]{Iman Lashgari\corref{mycorrespondingauthor}}
\cortext[mycorrespondingauthor]{Corresponding author: imanl@mech.kth.se}

\author[address1,address2]{Francesco Picano}

\author[address3]{Wim Paul Breugem}

\author[address1]{Luca Brandt}

\address[address1]{Linn$\acute{\textrm{e}}$ Flow Centre and SeRC (Swedish e-Science Research Centre),KTH Mechanics, S-100 44 Stockholm, Sweden}
\address[address2]{Industrial Engineering Department, University of Padova, Padova, Italy}
\address[address3]{Laboratory for Aero $\&$ Hydrodynamics, TU-Delft, Delft, The Netherlands}

\begin{abstract}
We consider suspensions of neutrally-buoyant  finite-size rigid spherical particles in channel flow and investigate the relation between the particle dynamics and the mean bulk behavior of the mixture for Reynolds numbers $500 \le Re \le 5000$ and particle volume fraction $0\le \Phi \le 0.3$, via fully resolved numerical simulations. Analysis of the momentum balance reveals the existence of three different regimes: laminar, turbulent and inertial shear-thickening depending on which of the stress terms, viscous, Reynolds or particle stress, is the major responsible for the momentum transfer across the channel.
We show that both Reynolds and particle stress dominated flows fall into the Bagnoldian inertial regime and 
that the Bagnold number can predict the bulk behavior although this is due to two distinct physical mechanisms. 
A turbulent flow is characterized by larger particle dispersion and a more uniform particle distribution, whereas the particulate-dominated flows is 
associated with a significant particle migration towards the channel center where the flow is smooth laminar-like and dispersion low.
Interestingly, the collision kernel shows similar values in the different regimes, although the relative particle velocity and clustering clearly vary 
with inertia and particle concentration.
%


\end{abstract}


\end{frontmatter}


\section{Introduction}
Particles suspended in a carrier fluid can be found in many  biological, geophysical and  industrial flows. Some obvious examples are the blood flow in the human body, pyroclastic flows from volcanos, sedimentations in sea beds, fluidized beds and slurry flows.
Moreover, the knowledge of the particle dynamics is relevant,  among others, in biomechanical applications for extracorporeal devices and formation of clots.
Suspensions are typically employed to transport and mix particles by means of a carrier fluid \cite[][]{Eckstein77}. The overall effect of particles on the flow dynamics has therefore a significant impact on the energy consumption of biological and industrial processes.  Despite the numerous applications, however, it is still difficult to estimate the force needed to drive suspensions and the internal dissipation mechanisms are not fully understood, especially in a turbulent flow. Unlike single phase flows where the pressure drop can be accurately predicted  as a function of the Reynolds number and the {properties} of the wall surface (roughness effects), additional parameters become relevant in the presence of a suspended phase when the properties of the particles  (size, shape, density, stiffness, volume fraction, mass fraction) affect the overall dynamics of the suspension. The behavior of these multiphase flows becomes even more complicated when the particle volume fraction is high, inertial effects are non-negligible and particles have finite size, i.e. size of the order of the relevant flow structures \cite[][]{Campbell90}.

In this study we focus on non-colloidal suspensions, mixtures where the dispersed particles are greater than colloidal in size and thermal fluctuations are negligible. As Brownian motion is negligible there is no diffusion to create an equilibrium structure making the problem one of fundamental
non-equilibrium physics. The aim of this study is to gain physical understanding of  the role of the fluctuations induced by the suspended phase and their coupling to the mean flow, the effect of particle inertia and the modifications of the particle interactions when increasing the {(bulk flow) Reynolds number}. As shown also here, it is fundamental to examine the local particle concentration, migration and segregation for a full comprehension of the transport processes at work. Inhomogeneities in the particle distribution are documented at low and finite Reynolds numbers, e.g.\ the so-called Segre-Silberberg effect \cite[][]{Segre61}. Here we document how the interactions between the turbulent flow structures and particle-induced disturbances alter the macroscopic flow behaviour.

Only few studies have been devoted to the inertial flow of suspensions in the presence of finite size particles. 
Matas \emph{et al.}\cite{Matas03} performed experiments with a suspension of neutrally buoyant particles in pipe flow and defined
the laminar and turbulent regimes according to the spectra of the pressure fluctuations between the inlet and exit of the pipe. The critical Reynolds number separating the existence of the two regimes exhibits a non-monotonic behaviour with the volume fractions for large enough particles. 
A result partially reproduced by the numerical simulations in Yu \emph{et al.}\cite{Yu13}. Since velocity fluctuations exist at all Reynolds numbers, these authors  choose the streamwise velocity perturbation kinetic energy as the criterion to distinguish between laminar and turbulent flow. A more detailed study on the transition of finite-size particle suspensions is performed by Loisel \emph{et. al.} \cite{Loisel13} for a fixed  volume fraction of about $5\%$.  The observed reduction of the critical Reynolds number is explained by the breakdown of the coherent flow structures to smaller and more energetic eddies, which prevents the flow re-laminarization when decreasing the Reynolds number. The characteristics of a fully turbulent channel flow laden with finite-size particles are presented in \cite{Picano15}, {such as the decrease in the Von Karman constant with increasing volume fraction and the increase in the overall drag}. 

The present work extends the analysis of Lashgari \emph{et al.}\cite{Lashgari14} on the inertial flow of suspensions of finite-size neutrally buoyant spherical particles. In this previous study, we document the existence of three different regimes when varying Reynolds number, $Re$, and particle volume fraction, $\Phi$. A laminar-like regime where viscous stress exhibits the strongest contribution to the total stress, a turbulent-like regime where the turbulent Reynolds stress mainly determines the momentum transfer across the channel (see also \cite{Picano15})
and a third regime, denoted as inertial shear-thickening, characterized by a significant enhancement of the wall shear stress that is not due to an increment of the Reynolds stress but due to the strong contribution of the particle stress. In the present work, we move our attention from the bulk flow behavior to the local behavior  by studying in detail the particle  dynamics, single and pair particle statistics. In particular, we examine the particle local volume fraction, dispersion coefficients  and collision kernels for the three regimes introduced in Lashgari \emph{et. al.} \cite{Lashgari14}. 
Our dataset is based on fully resolved numerical simulations of the two-phase system.

\begin{figure}
\centering{
\includegraphics[width=0.9\linewidth]{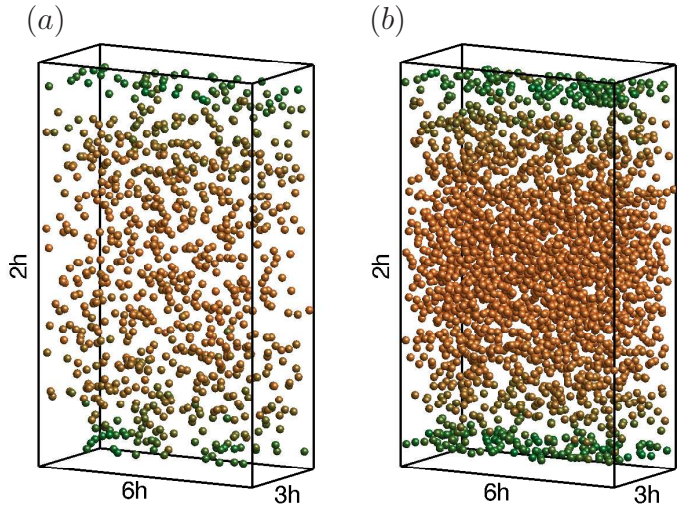}
\put(-145,200){{\large $(b)$}}
\put(-275,200){{\large $(a)$}}
\caption{\label{fig:vis} 
The instantaneous particle  arrangement for (a) a turbulent-like flow, $Re=5000$ $\&$ $\Phi=0.1$, and (b) a particle-stress dominated flow, $Re=2500$ $\&$ $\Phi=0.3$. The data are not visualized at true scale.}}
\end{figure}

We aim to connect our results to the seminal work by Bagnold \cite{Bagnold54}. Using experimental data of a suspension of neutrally 
buoyant solid particles in an annular domain between two concentric cylinders, Bagnold understood that the shearing of closely spaced particles would generate a normal or dispersive stress  in addition to the shear stress \cite[][]{Hunt02}.  He used the ratio between the grain inertia and the viscous stress to define different flow regimes. The viscous and inertial regimes introduced by Bagnold are characterized by a linear and quadratic relation between the wall shear/normal stress and the shear rate, {respectively}. Inspired by Bagnold's experiment, Fall \emph{et al.}\cite{Fall10} performed a similar study in plane Couette flow; these authors show a smooth transition from the Newtonian (viscous) to the Bagnoldian (inertial) regime by increasing the shear-rate. The laminar flow at high volume fractions behaves similarly to dry granular flows \cite[][]{Campbell90}: the flow experiences discontinuous shear-thickening and fast particle migration toward the regions of low shear. Both effects {(shear-thickening and particle migration towards region with low shear)} have been observed in several previous investigations of dense suspensions at low Reynolds number, see \cite{Hampton97,Brown09,Maxey11} among others. 
Shear-thickening at higher volume fractions is examined among others in Haddadi and Morris \cite{Morris14} who clearly identify the role of friction among particles in relative motion. 
The origin of shear-thickening in the presence of non-negligible inertia is attributed to the particle dynamic excluded volume in the recent work by Picano \emph{et al.}\cite{Picano13} at lower volume fractions.  The effective volume fraction of the suspension increases because of the shadow region (a region with statistically vanishing relative particle flux) around the particles. 
Particle migration across the channel is not an inertial effect and is observed also in Stokes flow at high volume fractions  \cite[][]{Maxey11}. The particles tend to migrate from regions of high to low shear due to the imbalance of the normal stress resulting from the particle interactions \cite[][]{guamor_book}. 

Less is known of the inertial Bagnoldian regime. It is worth mentioning that, for the same bulk behavior, the Bagnoldian regime can be either Reynolds stress or particle stress dominated, as deduced from the data in \cite{Lashgari14}.
{This finding motivated the present study where we focus on the particle dynamics to understand the two different underlying physical mechanisms.}

Understanding the dynamics of particle dispersion and collisions, especially when the particle inertia is non-negligible and the suspension is not dilute, is therefore important  due to their direct connection to the flow bulk properties, as also demonstrated in this study.
 The mutual and hydrodynamic interactions between the particles produce irregular motions, promote lateral migration from the instantaneous average particle trajectories and induce dispersion (for more details see \cite{Eckstein77,Breedveld98,Sierou04}). 
As an example, we report in figure~\ref{fig:vis} the instantaneous particle distribution for two different regimes:  i) a turbulent flow where transport is mainly determined by the Reynolds stresses and ii) a shear-thickening flow dominated by the particle stress.
Note that the wall normal direction is amplified by a factor 5 for the sake of clarity and the particle colors represent the magnitude of their translational velocities. We note an  uniform concentration for the turbulent-like flow (left panel) and an accumulation towards the channel centre for the flow dominated by the particle stress (right panel) that will be quantified and analyzed in this paper. 
 
Particle collisions are also relevant to the total momentum transfer and can be estimated from the relative position and velocity of the particle pairs \cite[][]{Collins97}: these can be directly connected to the particle diffusivity in the cross-stream direction and to accumulation  in specific regions \cite[][]{Hinch96,Collins00,Vincent12}. 
{The opposite is true for  Brownianan  suspensions where the particle concentration variation arises from gradient-induced diffusivity\cite{Breedveld98}, and finite-size effects are less important.} 
In this work we show a strong shear-induced self-diffusivity at high particle volume fractions which is not dependent on the Reynolds number and plays an important role in the collision dynamics and eventually on the bulk flow behavior.

This paper is organized as follows. We discuss the governing equations, the numerical method and validations in $\S2$. The  results of the simulations are discussed in  $\S3$, whereas conclusions and final remarks are presented  in $\S4$. 

\section{Governing equations and numerical method}

\subsection{Governing equations}
We study the motion of suspended rigid neutrally buoyant particles in a Newtonian carrier fluid. 
The Navier-Stokes and continuity equations govern the motion of the fluid  phase,
\begin{align}
\rho^f (\frac{\partial \textbf{u}}{\partial t} + \textbf{u} \cdot \nabla \textbf{u}) = -\nabla P + \mu^{f} \nabla^2 \textbf{u} + \rho^f \textbf{f}, \\ \nonumber
\nabla \cdot \textbf{u} = 0,
\label{eq:NS}  
\end{align}
where $\rho^{f}$ and $\mu^{f}$ are the density and viscosity of the fluid. 
Here we denote the spanwise,  streamwise and wall-normal coordinates as $(x,y,z)$ with corresponding {velocities $\textbf{u}=(u,v,w)$}.  The force on the fluid, $\textbf{f}$, is due to the presence of the finite-size particles.     
The motion of the particles is governed by the Newton-Euler equations   
\begin{align}
m^p \frac{ d \textbf{U}_c^{p}}{dt} = \textbf{F}^p , \\ \nonumber
I^p \frac{ d \pmb{\Omega}_c^{p}}{dt} = \textbf{T}^p , 
\label{eq:NS}  
\end{align}
where $m^p$ and $I^p$ are the mass and momentum inertia of particle $p$, $\textbf{U}_c^p$ and $\pmb{\Omega}_c^p$ the velocity and rotation  rate  and $\textbf{F}^p$ and $\textbf{T}^p$ the net force and momentum resulting from hydrodynamic and particle-particle interactions that for neutrally buoyant particles read
\begin{align}
F^p = \oint_{\partial {V}_p}  [ -PI + \mu^f (\nabla \textbf{u} + \nabla \textbf{u}^T ) ] \cdot  \textbf{n} dS+ \textbf{F}_c, \\ \nonumber
T^p = \oint_{\partial {V}_p}  \textbf{r} \times \big{\{} [ -PI + \mu^f (\nabla \textbf{u} + \nabla \textbf{u}^T ) ] \cdot  \textbf{n}  \big{\}} dS + \textbf{T}_c. 
\label{eq:NS}  
\end{align}
In this expression $\partial {V}_p$ represents the surface of the particles with unit normal vector $\textbf{n}$. The radial distance from the centre of the particle is denoted by $\textbf{r}$ and the force and torque resulting from particle-particle (particle-wall) contacts are indicated by  $\textbf{F}_c$ and $\textbf{T}_c$. 
The equations for the fluid and particle phase are coupled by the no slip and no penetration conditions on each point \textbf{X} on the surface of a particle, i.e. $\textbf{u}(\textbf{X})= \textbf{U}^p (\textbf{X}) = \textbf{U}_c^p + \pmb{\Omega}_c^p \times \textbf{r} $. 
We use here an Immersed Boundary Method \cite[][]{Breugem12}, where this condition is satisfied indirectly by applying the forcing $\textbf{f}$ on the right hand side of the Navier-Stokes equations.

\subsection{Numerical method}
We employ a Navier-Stokes solver coupled with an Immersed Boundary Method (IBM) to follow 
the motion of the fluid and rigid spheres in the domain. 
The direct forcing method was originally proposed by Uhlmann \cite{Uhlmann05} and {modified} by Breugem \cite{Breugem12} to ensure second-order spatial accuracy. An Eulerian fixed mesh is used for the fluid phase and a Lagrangian mesh to represent the moving surface of the particles. 
The IBM forcing imposes no-slip and no-penetration boundary conditions on the surface of the particles. 
When the distance between particles or with a wall becomes smaller or of the order of the mesh size, the interactions between the particles include  an additional lubrication correction. {Surface roughness effects are accounted for at very close approach. Finally, a soft-sphere collision model is employed to model collisions/contacts from the relative velocity and (slight) overlap of colliding particles.} (see the appendix of  \cite{Lambert13} for more details). 
The IBM code has been used to study passive and active suspensions by Lambert \emph{et al.}\cite{Lambert13} and Picano \emph{et al.}\cite{Picano13,Picano15}. 

\subsubsection{Validation}

\begin{figure}
\centering{
\includegraphics[trim = 10mm 20mm 10mm 25mm,width=0.6\linewidth]{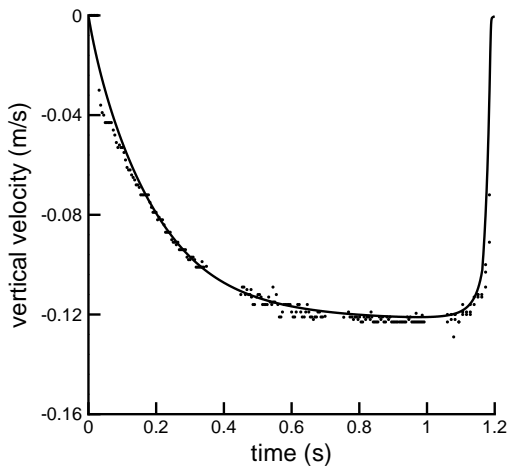}
\caption{\label{fig:Experiment-validation} 
{Particle settling velocity versus observation time; the solid line represents the numerical results whereas the dots are the experimental data from \cite{Cate02}}}}
\end{figure}

The accuracy of the code has been verified against several test cases in \cite{Breugem12}. In particular, the benchmark cases include:\\
1) Flow through a regular array of spheres at solid volume fraction of $6.5\%$ for which $2^{nd}$ order accuracy has been shown together with excellent agreement with theoretical predictions in \cite[][]{Hasimoto59}.\\
2) Lubrication force between 2 spheres: accurate predictions up to a gap width of $2.5\%$ of the sphere radius at resolution $D/dx=32$ (see \cite{Breugem10} and accuracy test at 1 specific gap width in \cite{Breugem12}).\\
3) Drafting-kissing-tumbling interaction between two spheres for which $2^{nd}$ order accuracy is confirmed.\\
4) Validations of the global suspension behavior are reported in  \cite{Picano13,Picano15}.\\
5) {The code has also been compared with the experimental data by Cate \emph{et. al.} \cite{Cate02} on a falling spherical particle in a closed rectangular container. The container dimensions were as in the experiment: $100 \times 100 \times 160$ mm in the 2 horizontal and the vertical direction, respectively. The diameter of the nylon sphere was 15 mm. The mass density of the nylon sphere 1120 kg/m3. The mass density of the working fluid, silicon oil in this case, was 960 kg/m3 (so the mass density ratio = 1.17). The dynamic viscosity of the working fluid is 0.058 kg/(ms) for estimated Reynolds number $Re \approx 31.9$. Figure~\ref{fig:Experiment-validation} shows the particle velocity as a function of time; the dots are the experimental data. The small difference at early time is probably related to the different release procedure in simulations and experiments.} 

\begin{figure}
\centering{
\includegraphics[width=0.7\linewidth]{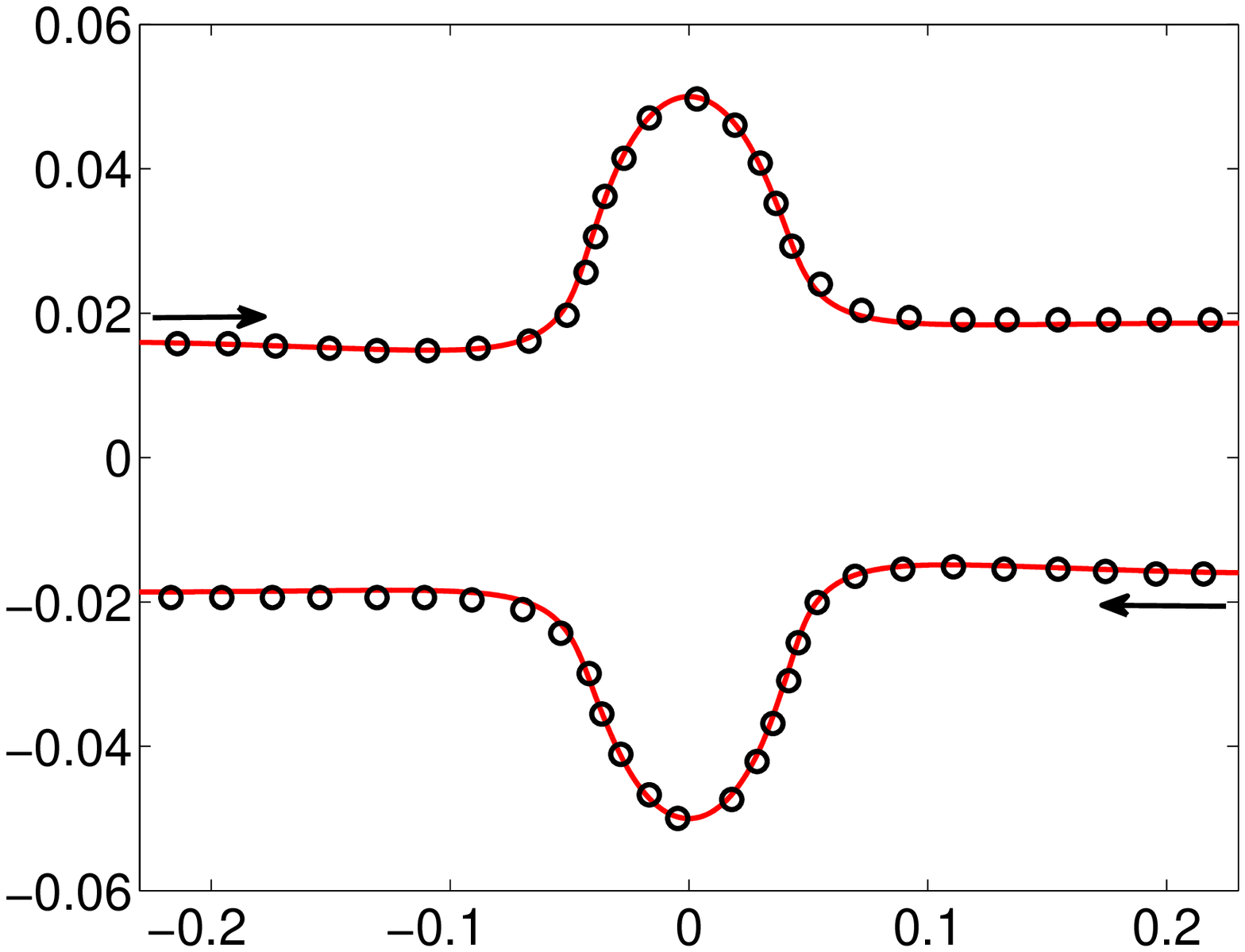}
\put(-245,90){{\large $z$}}
\put(-120,0){{\large $x$}}
\caption{\label{fig:validation_morris} Trajectory of a particle pair released in a simple shear flow from initial positions $\textbf{X}_1=(0,-4.85a,0.32a)$ and  $\textbf{X}_2=(0,+4.85a,-0.32a)$. The particle radius $a=0.05$. The solid line indicates our simulation while open circles are the reference data in \cite{Morrisjfm08}. 
The arrows show the initial direction of the particle velocities.}}
\end{figure}

In this work, we further validate the code by comparing the trajectory of a particle pair in homogeneous shear flow against the work of Kulkarni and Morris  \cite{Morrisjfm08}. These authors employ a Lattice-Boltzmann method to study the effect of particle inertia in a laminar flow. We choose the same box size of $20a\times 20a \times 20a$ with $a$ the particle radius and place the origin of the coordinate system at the center of the box. The two particles are initially positioned at $\textbf{X}_1=(0,-4.85a,0.32a)$ and  $\textbf{X}_2=(0,+4.85a,-0.32a)$ and move in opposite direction. The particle Reynolds number $Re_p= \frac{\dot\gamma a^2}{\nu}= 0.1$ with $\dot \gamma$ the imposed shear rate and $\nu$ the fluid kinematic viscosity. The initial particle velocity is the same as the local fluid velocity at the center of the particle and the initial rotation is equal to half the local fluid vorticity. The trajectory of the particle centers are displayed in figure \ref{fig:validation_morris} together with the data from \cite{Morrisjfm08} (open circles). The arrows display the direction of the particle velocity at the initial time.  The results show a good agreement: the particle trajectories deviate away from the centerline after their interaction (a typical inertial effect).

\subsection{Flow configuration and numerical setup}
We simulate a channel flow with periodic boundary conditions in both streamwise and spanwise directions. The box-size is $6h\times2h\times3h$ in the streamwise, wall-normal and spanwise directions. The particles have all the same radius, $a= h/10$. 
The Reynolds number is defined as $Re=\frac{2hU_b}{\nu}$, where $U_b$ and $\nu$ are the fluid bulk velocity and kinematic viscosity. A wide range of parameters have been considered; the Reynolds number $500 \le Re \le 5000$, and the particle volume fraction $0 \le \Phi \le 0.3$. The simulations start with a high amplitude localized disturbance in the form of two counter-rotating streamwise vorticities \cite[][]{Henningson91} to efficiently trigger turbulence, if the Reynolds number is high enough for it to be sustained \cite{Lashgari15}. The particles are randomly positioned at time zero, with velocity equal to the local fluid velocity and rotation equal to half the local fluid vorticity.  

{For the simulations presented in this work we employ  $480\times160\times240$ uniform Eulerian grid points in the streamwise, wall-normal and spanwise directions and 746 Lagrangian grid points to cover the surface of each particle.}{ The statistics are computed when the flow is fully developed. To check if the results are statistically converged, we repeat the analysis using half of the number of the samples and compare the outcome with the one from the total number of samples.}

\section{Results}
In this work we study the local properties of the particulate flow, in particular particle distributions, dispersions and collisions, and connect the results with the bulk flow regimes identified in \cite{Lashgari14}.


\subsection{Inertial regimes and Bagnold theory} 
In this section, we analyse in detail the momentum budget of the two phase flow and provide a comparison with the seminal work by Bagnold \cite{Bagnold54}.  Batchelor \cite{Batchelor70} was probably the first to derive an analytical expression for the bulk stress of suspensions of rigid particles and to discuss the relation between the macroscopic properties of a homogenous suspension and the flow structures at the particle scales. 
He assumed that i) the bulk stress depends on the instantaneous particles configuration in a flow element containing a large number of particles; ii) the configuration in each element depends on the history of the motion (memory effect). This shows the importance of the local microstructure and of its time history to determine the bulk behavior of the suspension. For colloidal suspensions in the inertialess regime, the relation between the particle and bulk scale structures are thoroughly reviewed by Morris \cite{mor_ra09}. 


We employ the phase-averaged momentum equations following the formulation developed in \cite{Marchioro99,Prosperetti04,Zhang10} where the effect of spatial non-uniformity over a finite scale larger than the particle size has been taken into account, unlike in the original formulation by Bachelor.  The phase average momentum equation on the volume ${\cal V}$ with boundary ${\cal S(V)}$  reads \cite[][]{Zhang10}
\begin{align}
\rho \int_{\cal V} (\xi \pmb{a}^p + (1-\xi) \pmb{a}^f)  \,d {\cal V} = 
 \oint_{\cal S(V)} [ \xi \pmb{\sigma}^p + (1-\xi) \pmb{\sigma}^f ] \cdot n \,d S, 
\end{align}
where $\xi$ is the phase indicator with values $\xi=0$ and $\xi=1$ for the fluid and particle phase. The stress and acceleration of the fluid and particle phase are denoted by $\pmb{\sigma}^f$, $\pmb{a}^f$, $\pmb{\sigma}^p$ and $\pmb{a}^p$ respectively. Assuming statistical homogeneity in the streamwise and spanwise directions, one can obtain an expression for the stress budget across the channel (see for the detail of the derivations the appendix of \cite{Picano15}) 
\begin{align}
\tau(z/h)= - \langle v'^{t}w'^{t}\rangle + \nu(1-\varphi)\frac{d V^f}{d z} + \frac{\varphi}{\rho}\langle \sigma^{p}_{yz}\rangle= 
\nu \frac{d V^f}{d z}\bigg{|}_w (1-\frac{z}{h}),
\label{eq:total_stress}
\end{align}
where $\tau(z/h)$ is the total stress as a function of wall normal coordinate, $z$, normalized by the channel half width, $h$. The first term in the stress budget is the total Reynolds stress, $\tau_R=\langle v'^{t}w'^{t}\rangle=(1-\varphi)\langle v'^{f} w'^{f} \rangle + \varphi \langle v'^{p} w'^{p}\rangle$, consisting of the fluid and particle Reynolds stress weighted by local particle volume fraction, $\varphi$. The second therm, $\tau_V=\nu(1-\varphi)\frac{d V^f}{d z}$, is the viscous stress whereas the third term, $\tau_P=\frac{\varphi}{\rho}\langle \sigma^{p}_{yz}\rangle$, is the stress due to the particles. The sum of the three terms is a linear function across the channel, as for a classic turbulent flow \cite[][]{Pope00}, with the wall shear stress $\tau_w= \nu \frac{d V^f}{d z}\big{|}_w$. 
The expression for the particle stress is discussed in detail in \cite{Batchelor70,Morrispof08, Zhang10}. Following \cite{Batchelor70}, the particle stress in the absence of an external torque reads
\begin{align} 
\sigma_{ij}^p = \frac{1}{V} \Sigma_{V} \int_{A_p} \frac{1}{2} \{ \sigma_{ik} x_j + \sigma_{jk} x_i \} n_k  dA  
- \frac{1}{V} \Sigma_{V} \int_{V_p} \frac{1}{2} \rho \{ f'_i x_j + f'_j x_i \} dV + \sigma_{ij}^c,
\end{align}
where $x$ is the material point, $f'_i$ is the local acceleration of the particle relative to the average acceleration, $A_p$ and $V_p$ are the particle surface area and volume  and $\sigma_{ij} = - P \delta_{ij} + \mu^f (\frac{\partial \textbf{u}_i}{\partial x_j} + \frac{\partial \textbf{u}_j}{\partial x_i})$ is the fluid stress tensor. The first term on the right hand side is the hydrodynamic stresslet resulting from the symmetric part of the shear stress. This is related to effective viscosity in dilute suspension \cite[][]{Prosperetti04}. The second term is the stress related to the particle acceleration and rotation with respect to the neighbouring flow and the last term is the inter-particle stress which depends on the near-field particle interactions and particle collisions.

\begin{figure}
\centering{
\hspace{-0.3cm}
\includegraphics[width=0.5\linewidth]{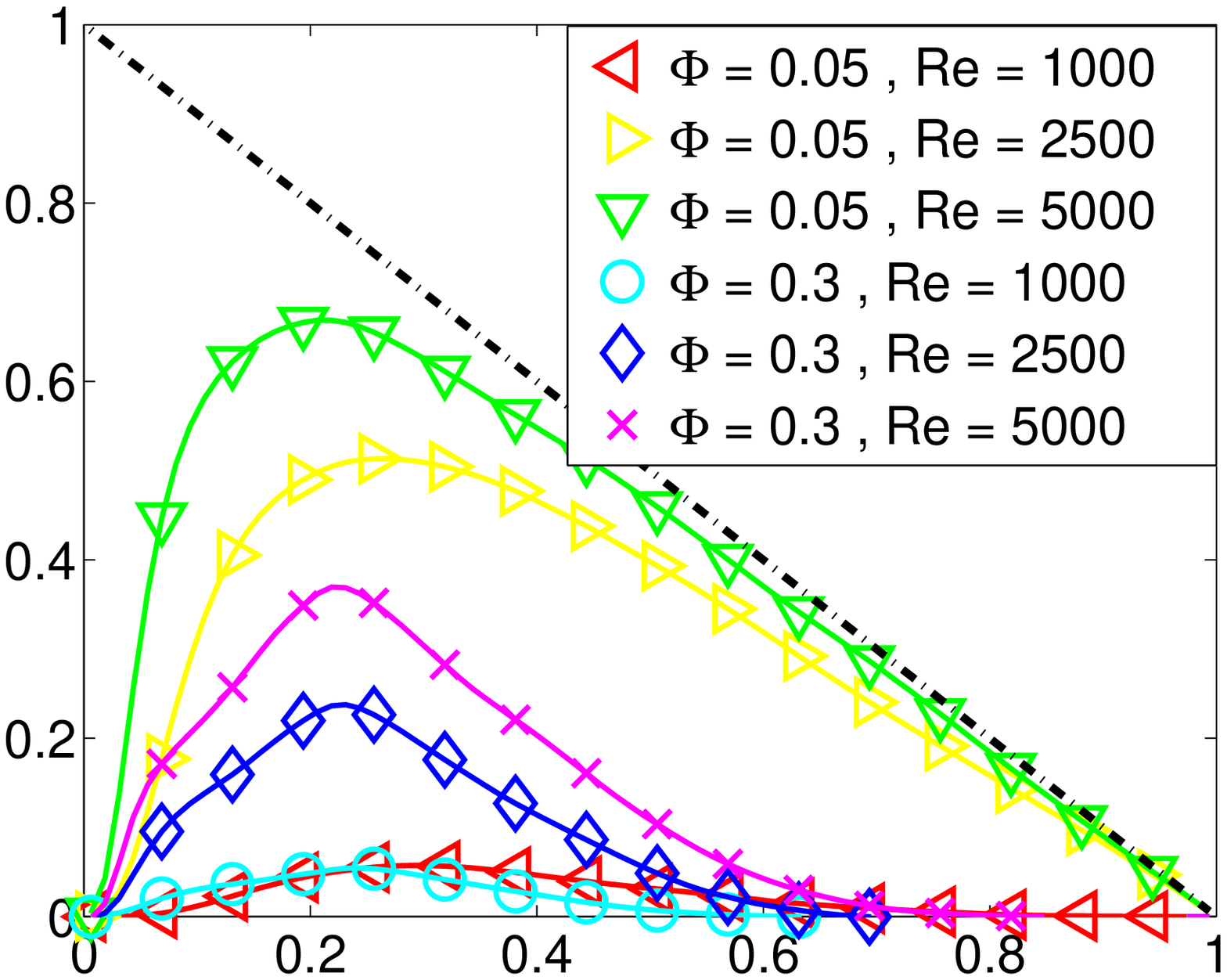}
\put(-170,125){{\large $(a)$}}
\put(-175,75){{$\tau_R$}}
\put(-90,0){{$z/h$}}
\includegraphics[width=0.5\linewidth]{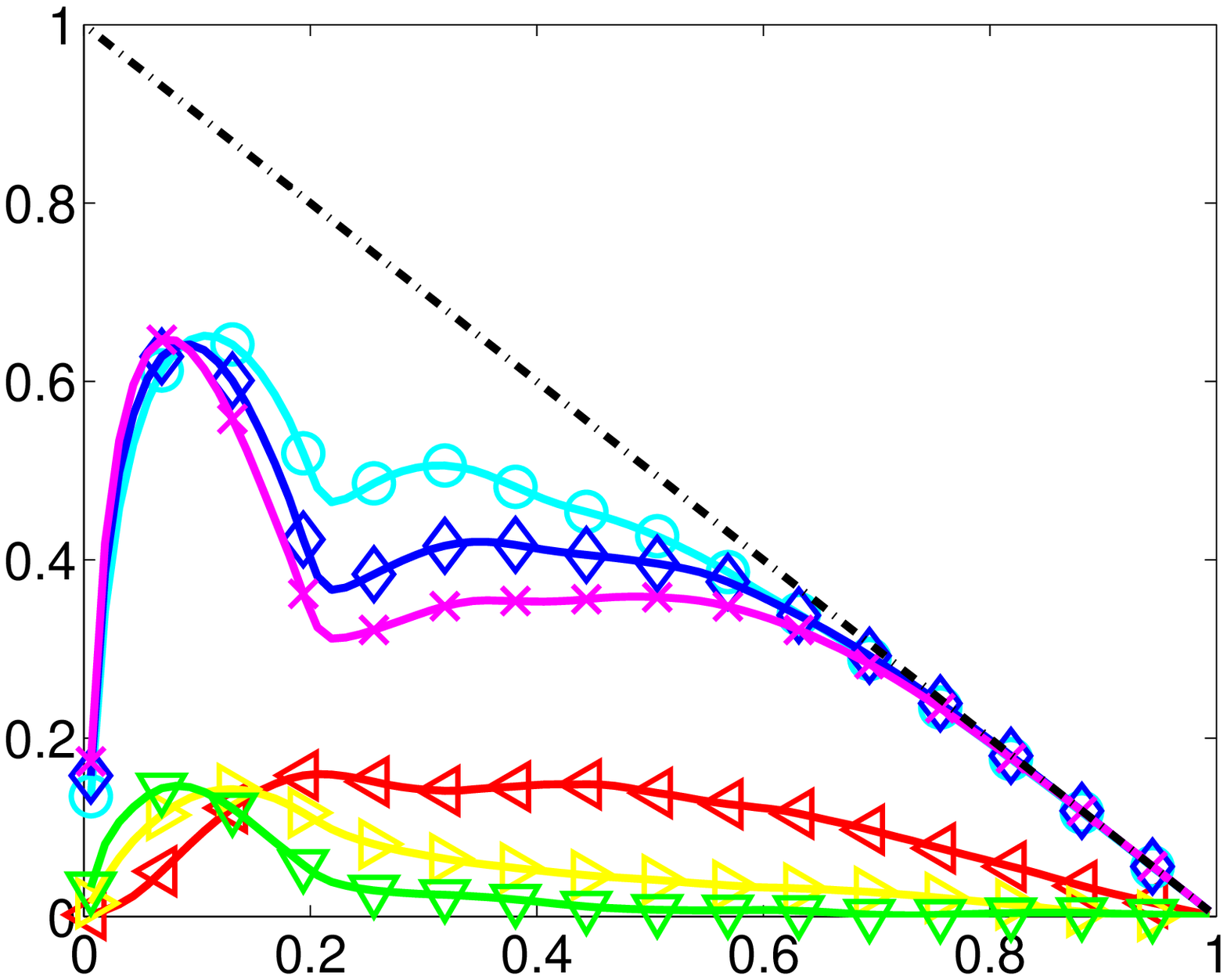}
\put(-170,125){{\large $(b)$}}
\put(-175,75){{$\tau_P$}}
\put(-90,-0){{$z/h$}}
\caption{\label{fig:stress-profile} 
Profiles of  (a) the Reynolds stress and (b) the particle stress across the channel for the cases indicated in the legend.}}
\end{figure}

The wall-normal profiles of Reynolds and particle stresses, $\tau_R$ and $\tau_P$, are displayed in figure \ref{fig:stress-profile} for flows at low and high particle volume fractions, $\Phi=0.05$ and $\Phi=0.3$. The cases at $\Phi=0.05$ and $Re>2500$ are characterized by a dominant contribution of the Reynolds stresses with significant particle stress only in the near-wall region. At high volume fractions, $\Phi=0.3$, on the contrary, the particle stress accounts for more  than $75\%$ of the total stress and the Reynolds stress is significant only in the intermediate region between the wall and centerline. As we will discuss in detail, the particle accumulation in the core region and in the layer close to the wall explains the high particle stress in the dense suspensions. Finally, we note that the major contribution to the total stress is due to the viscous forces at $\Phi=0.05$ and $Re=1000$; Increasing Reynolds number at fixed $\Phi$, the Reynolds stresses increases sharply when the flow becomes turbulent while the particle stress slightly decreases.

 \begin{figure}
\centering{
\hspace{-0.3cm}
\includegraphics[width=0.5\linewidth]{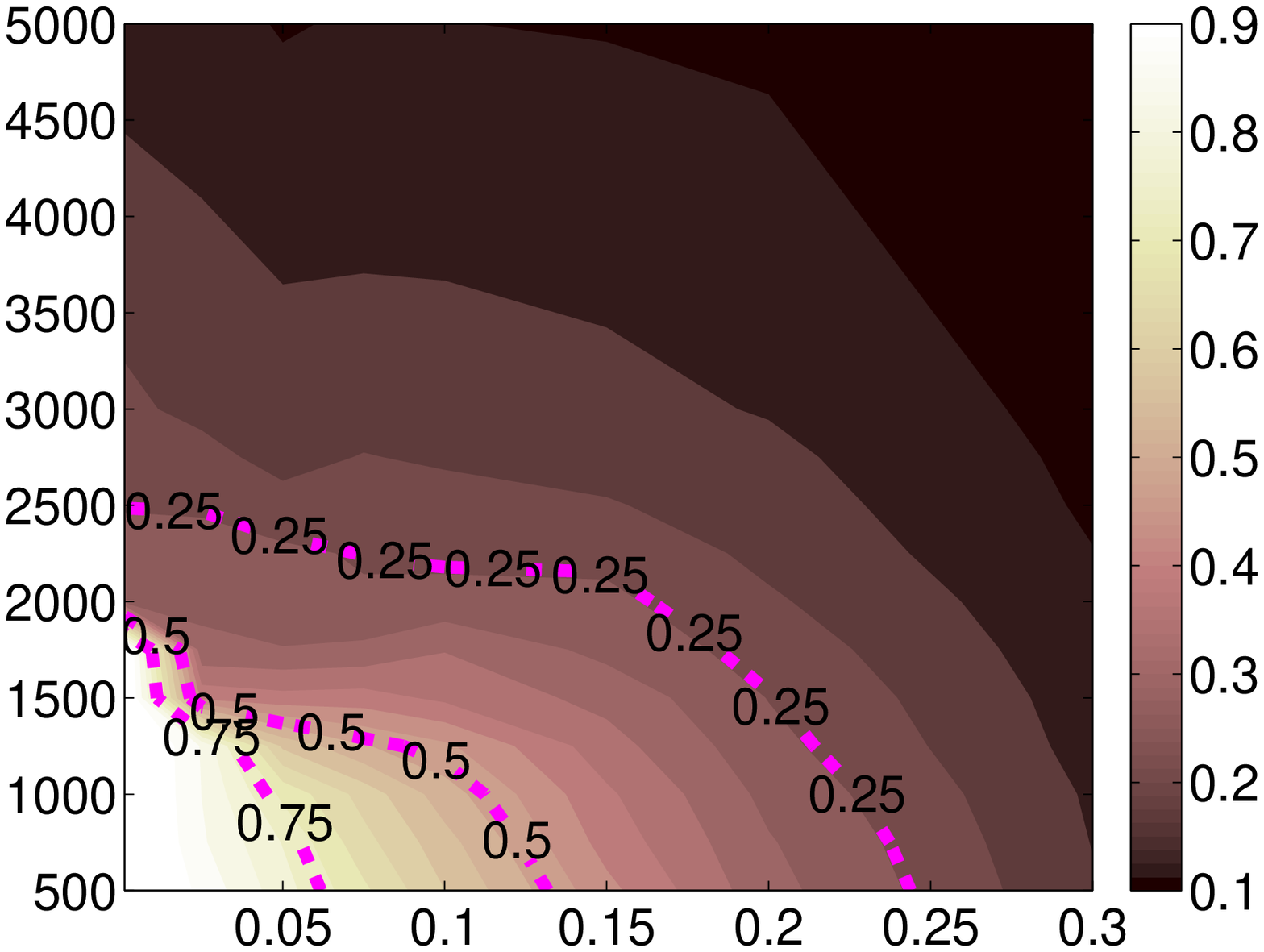}
\put(-160,130){{\large $(a)$}}
\put(-180,60){{$Re$}}
\put(-85,-5){{$\Phi$}}
\includegraphics[width=0.5\linewidth]{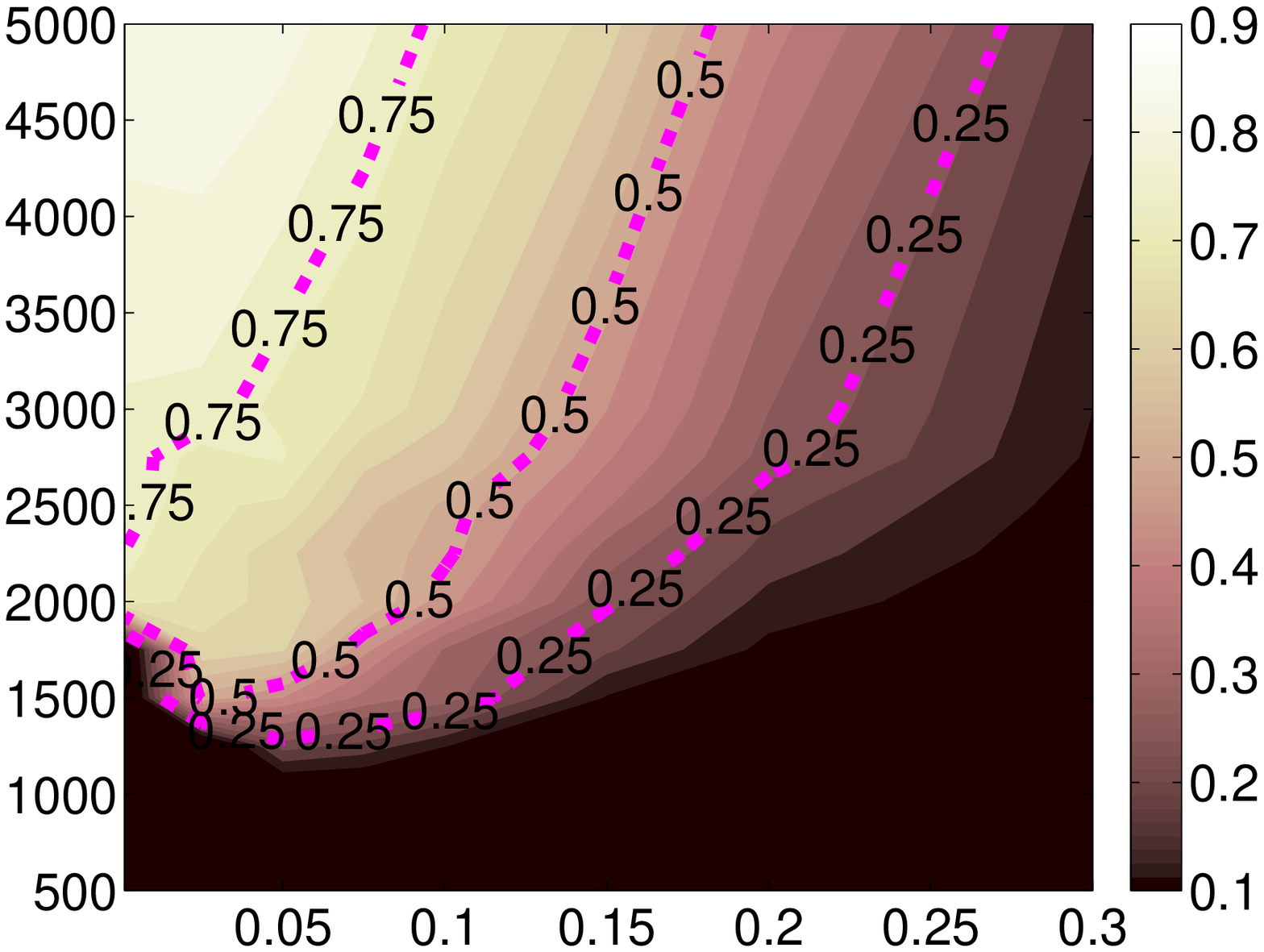}
\put(-160,130){{\large $(b)$}}
\put(-180,60){{$Re$}}
\put(-85,-5){{$\Phi$}}
\\
\includegraphics[width=0.5\linewidth]{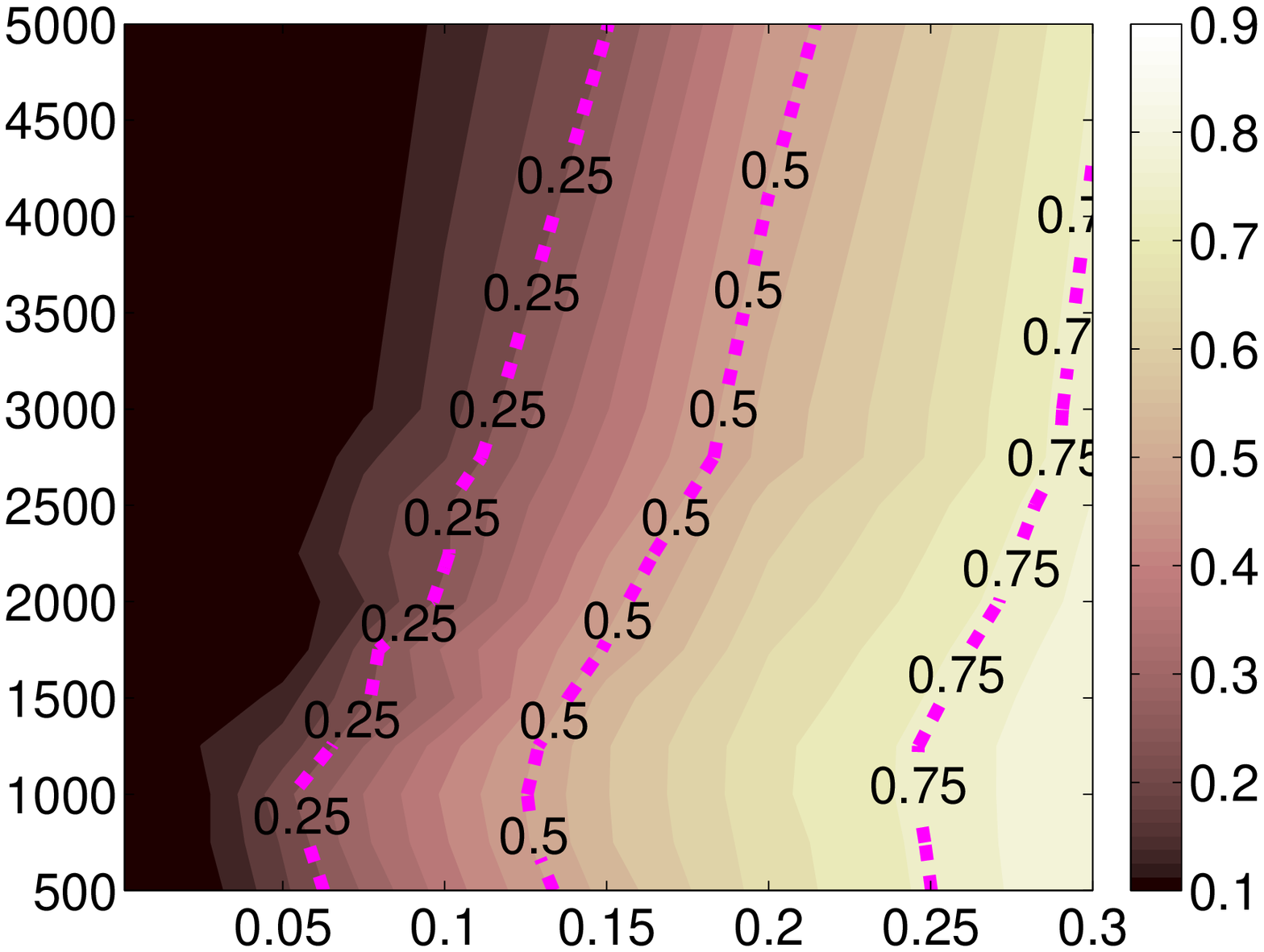}
\put(-160,130){{\large $(c)$}}
\put(-180,60){{$Re$}}
\put(-85,-5){{$\Phi$}}
\includegraphics[width=0.5\linewidth]{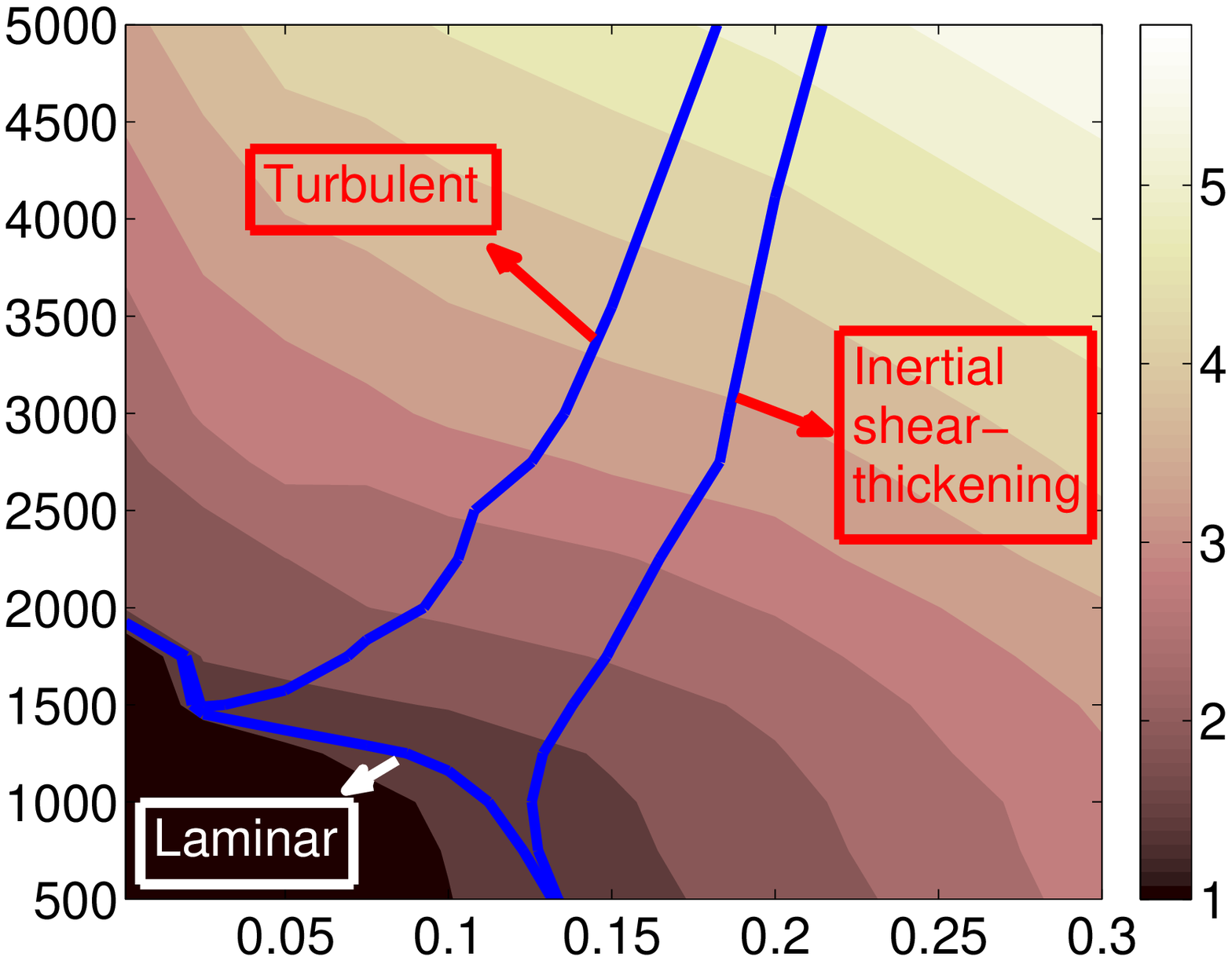}
\put(-160,130){{\large $(d)$}}
\put(-180,60){{$Re$}}
\put(-85,-5){{$\Phi$}}
\caption{\label{fig:maps} 
Contour map of the percentage contribution of (a) viscous stress, (b) Reynolds stress and (c) particle stress to the total momentum transport integrated across the channel. The isolines show the boundary of the regions in the map where the contribution of each term is more than $25\%$, $50\%$ and $75\%$. (d) Contour map of effective viscosity, the normalized wall shear stress divided by the shear at the wall of the corresponding laminar flow. }
}
\end{figure}   
To understand the role of the different transport mechanisms on the bulk flow behavior in the range of Reynolds numbers and particle volume fractions investigated, we show in figure \ref{fig:maps} (panels a,b,c) maps of the relative contribution of viscous, Reynolds and particle stress to the total momentum transfer integrated across the channel. The dashed lines represent iso-levels of $25\%$,  $50\%$ and $75\%$ of the total stress. The region where the viscous stress is more than $50\%$ of the total stress {is limited to $Re < 1900$ and $\Phi < 0.13$}. In this region, the action of viscous dissipation overcomes inertia. 
The contour lines in figure \ref{fig:maps}(b) show the non-monotonic behavior of the Reynolds stress which is also an indication of the level of fluctuations in the flow. This trend is in agreement with previous experimental and numerical findings \citep{Matas03,Yu13}, where the authors report a non-monotonic behavior of the critical conditions for the occurrence of turbulent flow when increasing the particle volume fraction. 
The contribution of the Reynolds stress  is more than $50\%$ of the total {for $Re>2000$ and $\Phi<0.1$}: the fluid and particle phases induce strong fluctuations that cannot be damped by viscous dissipation. 
%
The region with {$\Phi>0.13$} is characterized by values of  the particle stress larger than $50\%$ of the total stress. In this region, we expect a high level of hydrodynamic and particle-particle interactions that induce strong particle stresses. {Based on \ref{fig:maps}(b) and \ref{fig:maps}(c), the rate at which the particle stress contribution is increasing with Re is similar to the rate at which the Reynolds shear stress contribution is increasing (the lines have similar slopes). This suggests that for high $\Phi$ the flow will not be dominated by turbulent transport when increasing the Reynolds number \cite[][]{Lashgari14} }

Finally, we display the effective viscosity in figure~\ref{fig:maps}(d) . Following previous literature, we define the effective viscosity as the normalized wall shear stress divided by the shear at the wall of the corresponding laminar flow, $\mu_r= \tau_w / \tau_0$ \cite[][]{Cokelet99}.  We observe a monotonic increase of the dissipation when increasing both the Reynolds number and the particle volume fraction. 
The regions where the contribution of each stress term is more than 50 $\%$ of the total (See panels a,b and c) are depicted on the map of panel (d) by solid blue lines. Following Lashgari \emph{et al.}\cite{Lashgari14}, these regions  represent  the laminar, turbulent and inertial shear-thickening regimes where the viscous, Reynolds and particle stress contribute the most to the momentum transfer. The three transport mechanisms coexist with different relevance depending on the Reynolds number and particle volume fraction.

To continue our analysis, we first recall the work by Bagnold \cite{Bagnold54} on inertial suspensions. This author introduces what is known as the Bagnold number, {$Ba= 4 Re_p\sqrt{\lambda}$}, where $\lambda=\frac{1}{(0.74/\Phi)^{1/3}-1}$ is the linear concentration computed as the ratio between the particle diameter and the average radial separation distance and $Re_p$, the particle Reynolds number. This non-dimensional parameter represents the ratio between inertial and viscous stresses and it is shown to describe the bulk behavior of the suspensions reasonably well. Bagnold defines a macro-viscous regime, $Ba<40$, at low Reynolds number and low 
particle volume fraction where the relation between the stress and shear-rate is linear (similar to a Newtonian flow). The  Bagnoldian regime, instead, $Ba>450$, appears at higher Reynolds numbers and particle volume fractions and is characterized by a quadratic dependence of the stress on the shear-rate. 
\begin{figure}
\centering{
\hspace{-0.3cm}
\includegraphics[width=0.55\linewidth]{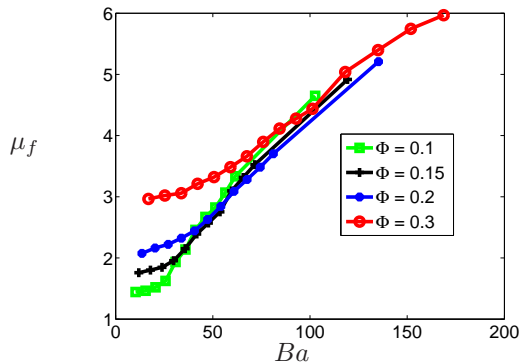}
\put(-205,80){{$\mu_f$}}
\put(-105,0){{$Ba$}}
\caption{\label{fig:mu-Ba} 
Effective viscosity, the normalized wall shear stress divided by the shear at the wall of the corresponding laminar flow, $\mu_r= \tau_w / \tau_0$, vs Bagnold number for four representative values of the volume fraction $\phi$. }}
\end{figure}

The effective viscosity pertaining our simulations is depicted versus the Bagnold number in figure \ref{fig:mu-Ba} for four representative values of the particle volume fraction. Interestingly, we observe that the effective viscosity of the suspension is almost constant when $Ba<40$, as expected in the visco-macro regime where the shear stress depends linearly on the Bagnold number (constant effective viscosity). 
When  $Ba>70$, all the curves collapse on a single line (see figure \ref{fig:maps}d) and the effective viscosity varies linearly with the Bagnold number.  

These results suggest that both the Reynolds and particle stress dominated flows fall into the Bagnoldian inertial regime: the same value of the Bagnold number and the same dissipation (effective viscosity) may therefore be explained by two different underlying physical mechanisms. The shear stress due to the residual turbulence becomes negligible when increasing the grain concentration, as predicted by Bagnold, and the particle stress takes the place of the Reynolds stress in the transport of momentum across the channel. At very high volume fraction the dynamics of the flow resemble granular media where the effect of interstitial fluid is negligible and the inter-particle collision is the main transport mechanism \cite[][]{Balachandar10}.


\subsection{Single particle statistics} 
The single particle statistics are computed by considering quantities related to each individual  particle and taking ensemble average over time and space. In particular, we extract the local volume fraction, mean and rms velocities of the particles as a function of the wall-normal coordinate $z$. 

\begin{figure}
\centering{
\includegraphics[width=0.6\linewidth]{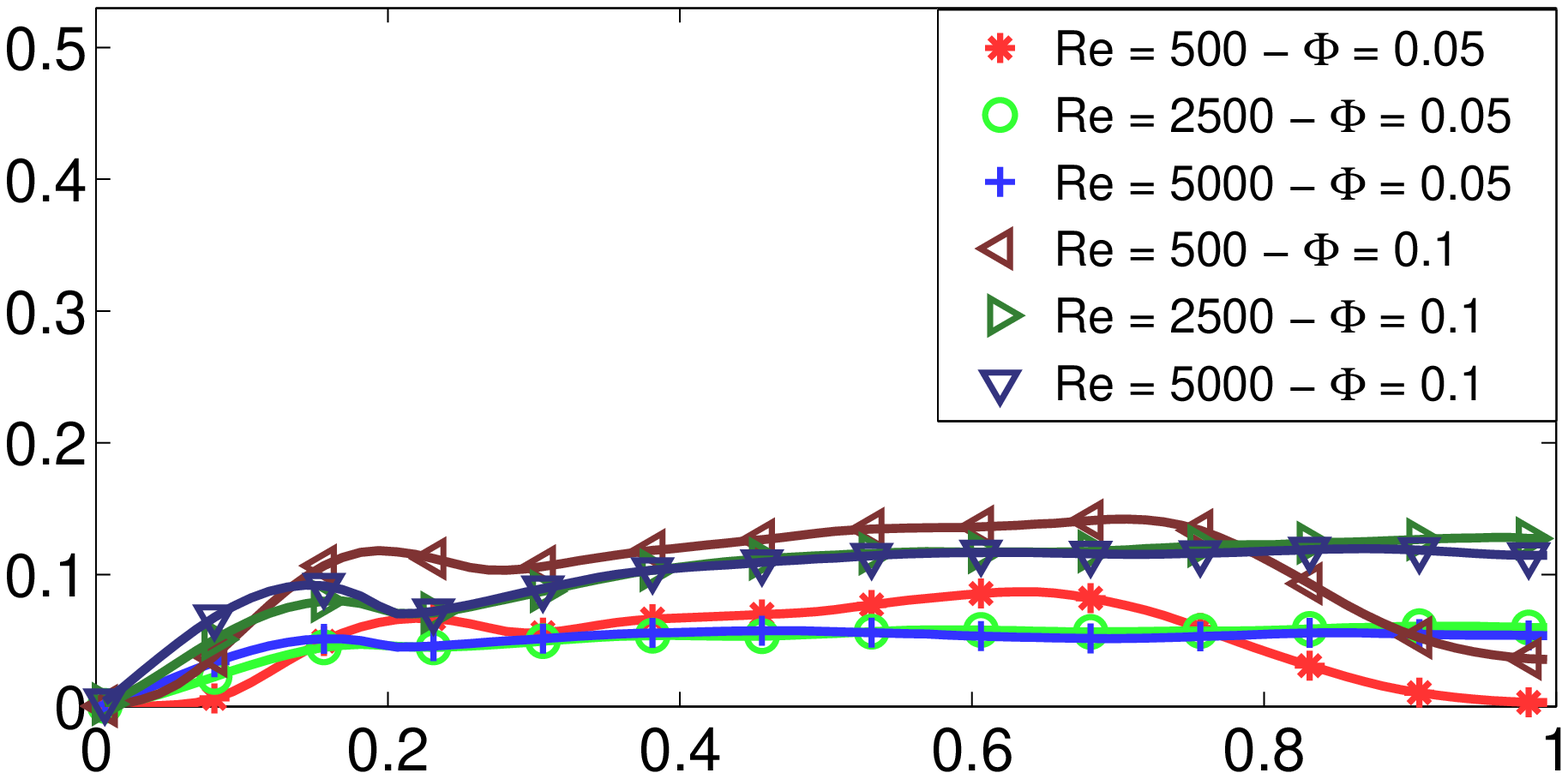}
\put(-215,65){{\large $\phi$}}\\
\includegraphics[width=0.6\linewidth]{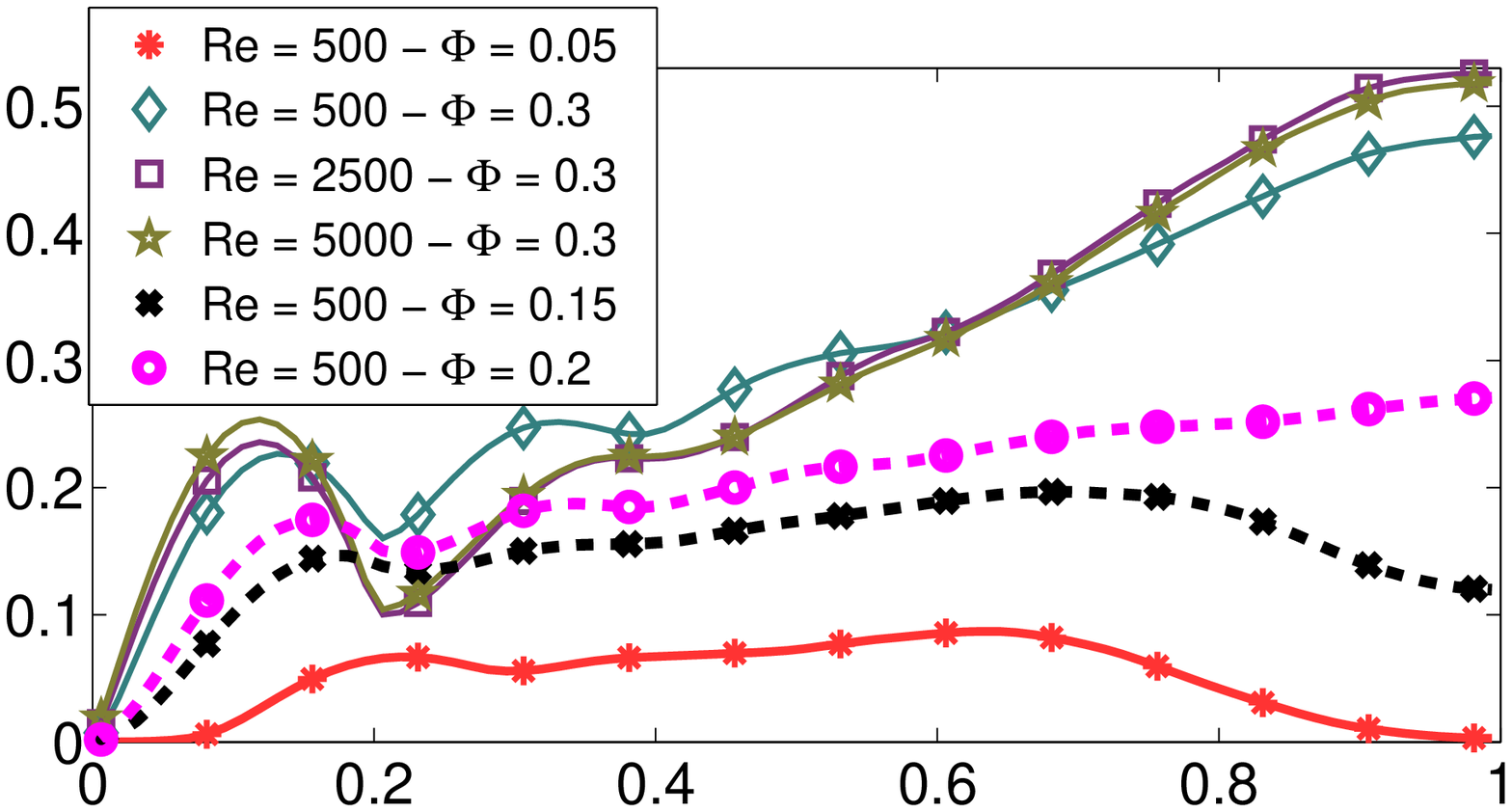}
\put(-215,65){{\large $\phi$}}
\put(-100,-5){{\large$z/h$}}
\caption{\label{fig:local_phi} 
Wall-normal profiles of the local volume fraction for different values of the Reynolds number $Re$ and volume fraction $\Phi$, see legend.}}
\end{figure}

The wall-normal profiles of the local particle volume fraction are shown in figure \ref{fig:local_phi} for different Reynolds numbers and particle volume fractions, covering the three different regimes introduced above. 
Based on the phase diagram in figure \ref{fig:maps}(d), we see that in the laminar regime 
the particles accumulate in the intermediate region between the wall and the channel centerline, $0.2 \lesssim z/h \lesssim 0.8$ ( cf. data for $Re=500$, $\Phi=0.05$ and $\Phi=0.1$).
This appears to be due to the Segre-Silberberg effect \cite[][]{Segre61}, an inertial effect (inertial migration) resulting from particle-fluid interactions, in particular explained in the dilute regime by the balance between the Saffman lift \cite[][]{Saffman65} and {inhomogeneous shear rate and wall effects (see paper by Schonberg and Hinch \cite{Schonberg89}).} The Segre-Silberberg effect is documented also at finite Reynolds numbers and volume fractions in the work by Matas \emph{et al.}\cite{Matas04} where it is shown that the particle equilibrium position moves closer to the wall when increasing the Reynolds number, i.e.\ {increasing inertial effects}. 

The data in the figure show that increasing the Reynolds number while keeping a low volume fraction the flow becomes turbulent and the particle distribution is almost uniform across the channel, except in the near wall region due to the one sided wall-particle interactions. 
The particle distribution is homogenized by the action of the Reynolds stresses for the flow cases at $Re=\{2500,5000\}$ and $\Phi=\{0.05 ,0.1\}$, which are in the turbulent region of the phase diagram in figure \ref{fig:maps}(d).

The particle distribution in the inertial shear-thickening regime, i.e.\ $\Phi=0.3$, exhibits a completely different behaviour with
a significant accumulation of the particles in the core region. 
The tendency of the particles to migrate toward the channel centreline is not {\it per se} an inertial effect (shear-induced migration) and is attributed to the imbalance of the normal stresses in the wall-normal direction \cite[see Ref.][for more details]{Fall10,Maxey11,guamor_book}. 
{When particle layers are sheared over each other, normal particle stress tends to push the particle layers further apart from each other which causes migration towards the core.}
Indeed, Notte and Brady \cite{Nott94} first documented particle migration towards the core region at $Re=0$. 
This effect is also evident from the data presented here as the profile of the local volume fraction does not change considerably increasing the Reynolds number from $500$ to $5000$.  The local particle volume fraction in the core region approaches {the value for a random loose packing}; here the particles experience an almost uniform translational velocity, of the same magnitude as the carrier flow in some sort of plug flow, see discussion below. 
 We find a peak of the local particle volume fraction at $z/h=0.1$ also at the highest volume fraction under investigation, corresponding to particle layering at the wall. Once a particle approaches the wall it tends to stay there because the interaction with neighbouring suspended particles is asymmetric and the strong near-wall lubrication force hinders departing motions.

We finally note that, as we increase the particle volume fraction above $\Phi=0.2$ at fixed Reynolds number, $Re=500$,  
the particle distribution changes from that typical of the Segre-Silberberg effect (due to the fluid-particle interactions) to display a significant accumulation in the core region due to the particle-particle interactions (shear-induced migration).

\begin{figure}
\centering{
\includegraphics[width=0.45\linewidth]{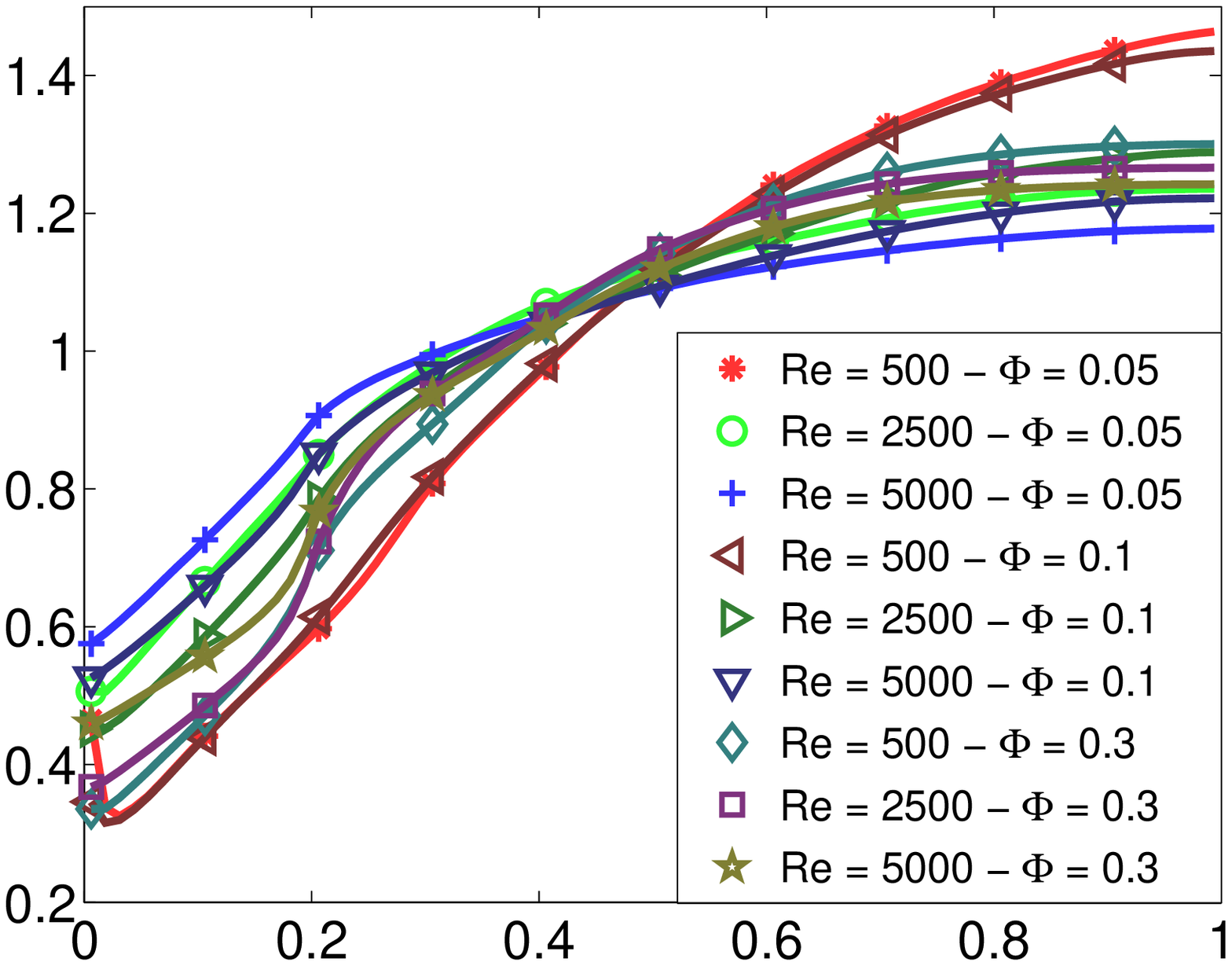}
\put(-155,115){{\large $(a)$}}
\put(-165,60){{$Vp$}}
\put(-85,0){{$z/h$}}
\includegraphics[width=0.45\linewidth]{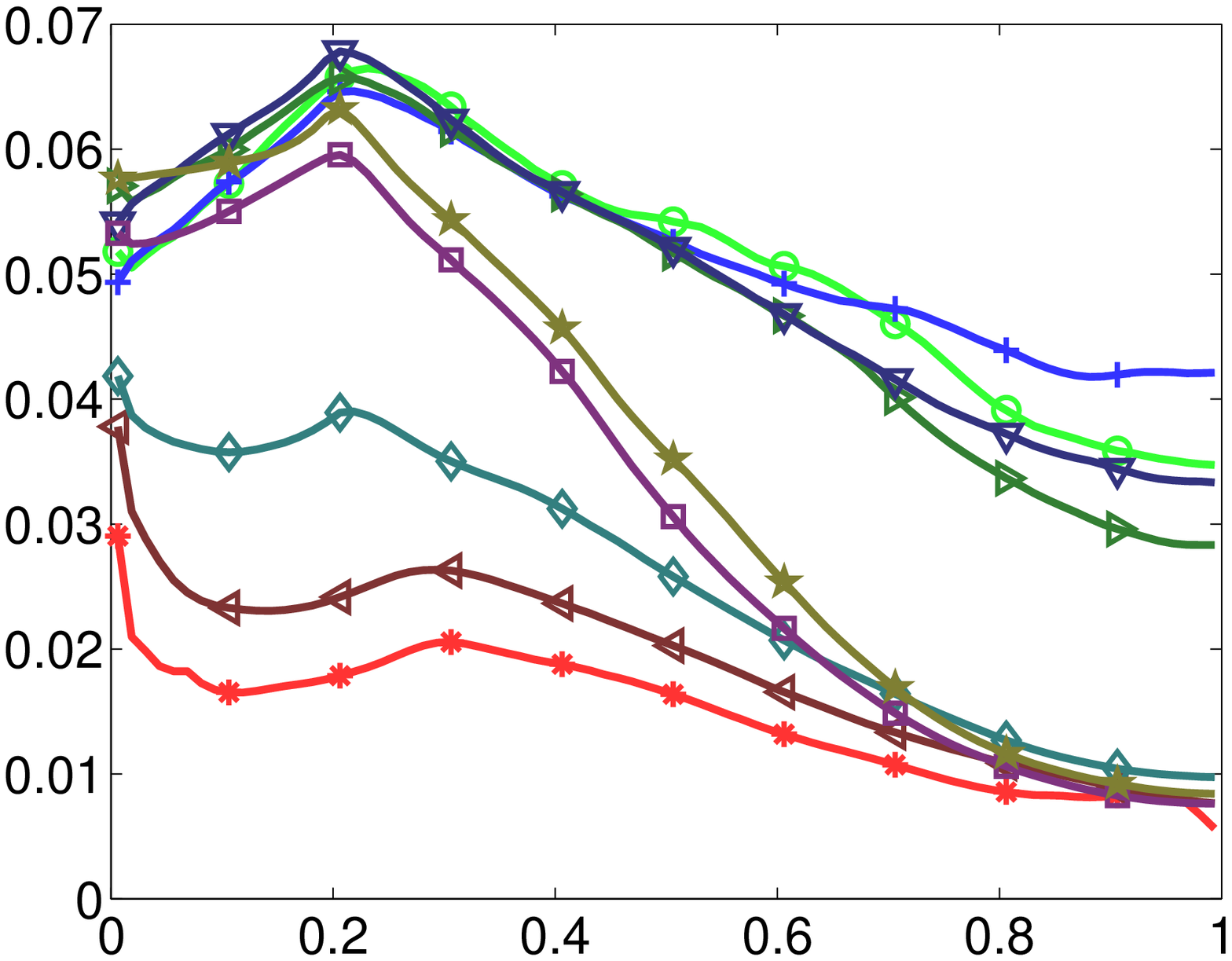}
\put(-155,115){{\large $(b)$}}
\put(-162,55){{\begin{turn}{90}$up'_{rms}$\end{turn}}}
\put(-85,0){{$z/h$}}
\\
\includegraphics[width=0.45\linewidth]{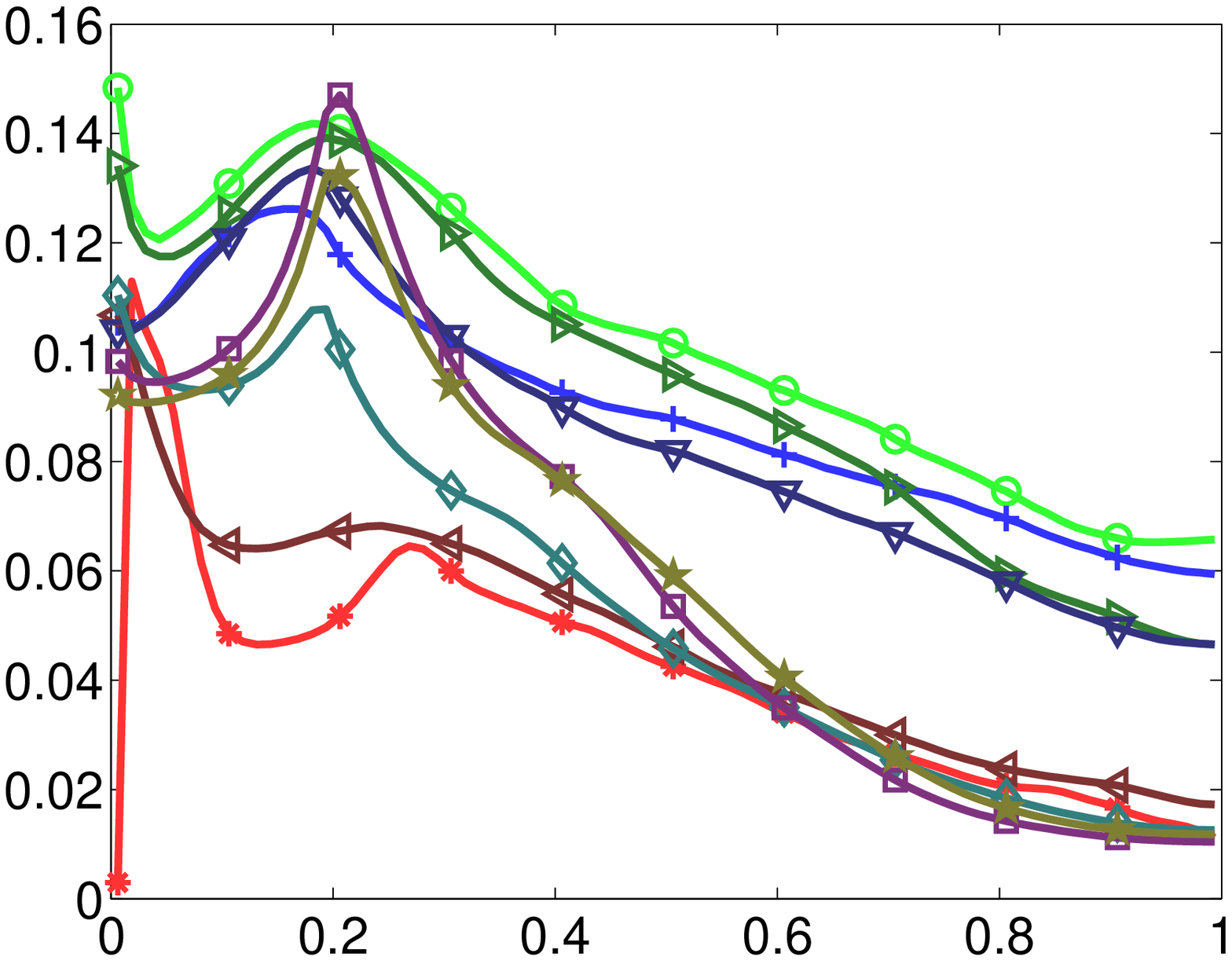}
\put(-155,115){{\large $(c)$}}
\put(-162,55){{\begin{turn}{90}$vp'_{rms}$\end{turn}}}
\put(-85,0){{$z/h$}}
\includegraphics[width=0.45\linewidth]{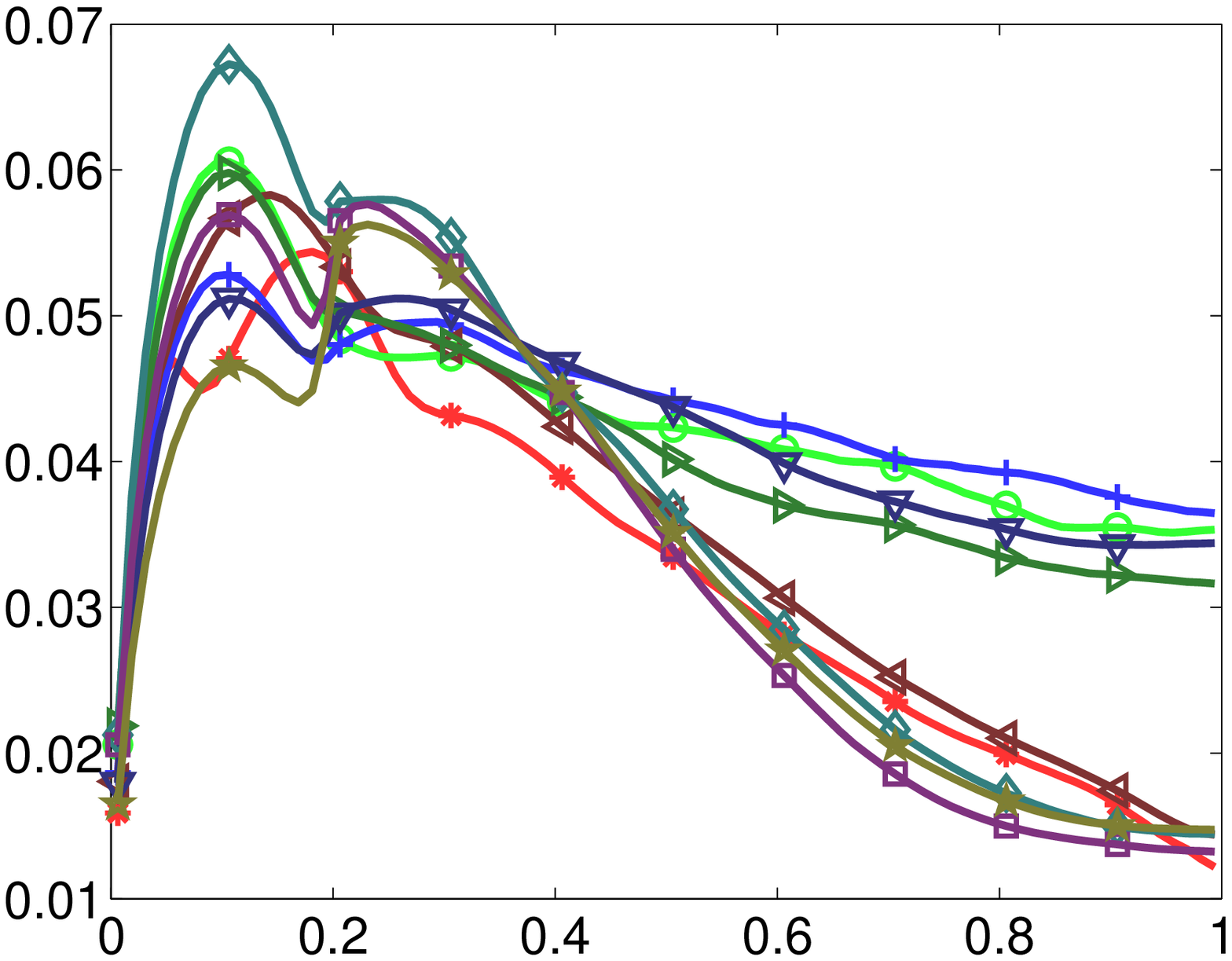}
\put(-155,115){{\large $(d)$}}
\put(-162,55){{\begin{turn}{90}$wp'_{rms}$\end{turn}}}
\put(-85,0){{$z/h$}}
\caption{\label{fig:par_rms} 
Wall-normal profiles of the mean particle streamwise velocity component and of the 3 particle r.m.s. velocities 
 for the different values of $Re$ and $\Phi$ indicated by the legend.}}
\end{figure}

The mean and rms particle velocities are depicted in figure \ref{fig:par_rms} for the same representative cases above. 
Figure \ref{fig:par_rms}(a) shows the mean streamwise velocity component  to highlight the slip velocities at the wall. The velocity profile is closer to the parabolic single-phase profile for the laminar cases, while it becomes blunt in the turbulent and inertial shear-thickening regimes. 
Increasing the Reynolds number, the mean flow becomes more uniform across the channel. The fluctuation (rms) velocities, figure \ref{fig:par_rms}(b,c,d), have also nonzero values at the wall: the level of fluctuations is higher for the turbulent cases and lower for the laminar flows, as expected. Interestingly, the level of particle velocity fluctuations of the inertial shear thickening regime {(in particular for the cases the cases $Re = 2500$, $\Phi=0.3$ and $Re = 5000$, $\Phi = 0.3$)} is similar to that of the turbulent flows closer to the wall,  $z/h<0.35$, and closer to the laminar values towards the centerline,  $z/h>0.65$. 
This observation suggests that the dynamics of the inertial shear-thickening flows are similar to that of the laminar flow in the core of the channel and to that of the turbulent flow in the near-wall region.
The low level of fluctuations in the core region results from the significant particle accumulation discussed above, which also explains the large particle stress in the inertial shear-thickening regime. 
Finally, we note that the peak of the wall-normal velocity fluctuations is located at $z/h=0.1$ for all the cases (except $Re=500$ and $\Phi=0.05$), indicating the relevance of the impact with the wall. The peak moves to $z/h=0.2$ for the streamwise and spanwise components of the velocity fluctuations.  

\subsection{Particle dispersion}

To understand the dynamics of the interactions in the flow,
we study the particle dispersion. 
The hydrodynamic and particle-particle interactions induce lateral forces and diffuse the particles away form their initial path. The dispersion is quantified by the variance of the particle displacement \cite[][]{Hinch96,Florian13}. Here, we compute the mean-square displacement of the particle trajectories, $\textbf{X}_p(t)=[x_p(t),y_p(t),z_p(t)]$, as a function of the time interval $\Delta t$,  averaging over all times, $t$, and particles, $p$, sampling after the initial transient in the flow development,   
\begin{align}
<\Delta \textbf{X}_p^2(\Delta t)>  =  < [ \textbf{X}_p ( t + \Delta t ) - \textbf{X}_p (t)]^2>_{p,t}.
\label{eq:mean_square_dis}
\end{align}
The sampling time for diffusion must be larger in the case of lower $\Phi$, since the fewer interactions between the particles require more time to reach the {statistical convergence} \cite[][]{Sierou04}.
The particle diffusion coefficients can then be calculated by measuring the slope of  the mean square displacement 
\begin{align}
\textbf{D} = \frac{<\Delta \textbf{X}_p^2(\Delta t)>}{2 \Delta t}, 
\label{eq:Dispersion}
\end{align}
where $\textbf{D}$ is the matrix containing the six independent diffusion coefficients $D_{ij}$. 
The normalized particle correlation is defined by    
\begin{align}
\textbf{R}(\Delta t) = \frac{< \textbf{X}_p(t + \Delta t)\textbf{X}_p(t )  >_{p,t} - < \textbf{X}_p> ^2}{< \textbf{X}_p ^2>- < \textbf{X}_p> ^2},
\label{eq:Correlation}
\end{align}
and it is used to quantify memory effects. The particle correlation and mean square displacement can be directly connected by 
\\
\begin{align}
\textbf{R}(\Delta t) = 1+ \frac{- \textbf{D} \Delta t}{< \textbf{X}_p ^2>- < \textbf{X}_p> ^2}.
\label{eq:Correlation_Dispersion}
\end{align}

\begin{figure}
\centering{
\includegraphics[width=1.0\linewidth]{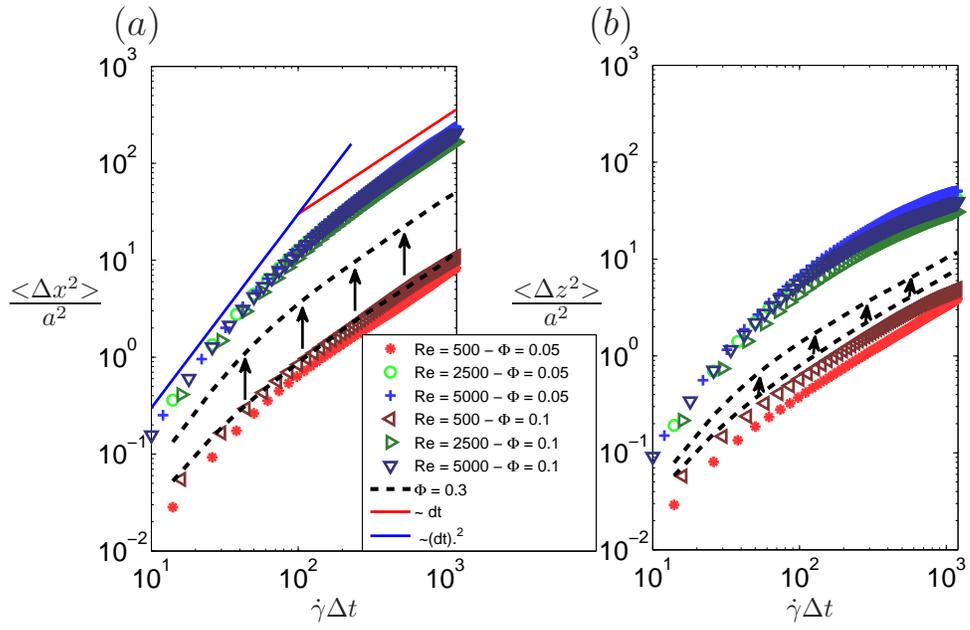}
\put(-180,115){{\Large $\frac{<\Delta z^2>}{a^2}$}}
\put(-370,115){{\Large $\frac{<\Delta x^2>}{a^2}$}}
\put(-75,0){{$\dot{\gamma} \Delta t$}}
\put(-255,0){{$\dot{\gamma} \Delta t$}}
\put(-330,220){{\Large $(a)$}}
\put(-150,220){{\Large $(b)$}}
\caption{\label{fig:Dispersion} 
Mean square particle displacement in the spanwise (a) and wall-normal direction (b) for the cases indicated in the legend. The thick black lines with arrows indicate the range of values assumed by flows in the inertial shear-thickening regime. }}
\end{figure}

The spanwise  mean square particle displacement, defined by eq.(\ref{eq:mean_square_dis}), is displayed in figure 
\ref{fig:Dispersion}(a) as a function of $\dot{\gamma} \Delta t$ where  $\dot{\gamma}=\frac{U_b}{h}$ is the average shear-rate across the channel. 
%
For all the cases studied, the mean 
square dispersion grows initially quadratically. In this ballistic regime, the particle trajectories are correlated and the displacements are proportional 
to $\Delta t $ so that $<\Delta x>^2 \propto (\Delta t)^2$. The trend changes for larger time lags when the classical diffusive behavior is retrieved. This 
is induced by particle-particle and hydrodynamic interactions that de-correlate the trajectories in time as shown among others by \cite{Sierou04} and  
\cite{Florian13}{in Stokes and low Reynolds number flows}. 
%
The asymptotic slope determines the dispersion coefficient:
The highest mean square displacement are found for $Re=2500-5000$ and $\Phi=0.05-0.1$ in the turbulent regime.
The strong fluctuations de-correlate the particles trajectories in shorter time and promote higher dispersion. The lowest dispersion 
is obtained in the simulations at $Re=500$ and $\Phi=0.05-0.1$, i.e. in the laminar regime where the viscous stress is dominating the flow dynamics, see figure \ref{fig:maps}. 

The asymptotic trend of the mean square displacement of the cases at $\Phi=0.3$ lies in between those of the turbulent and laminar regimes and smoothly increases with the Reynolds number. The two dashed black lines in the figure represent therefore the lower and upper limit attained at  $Re=500$ and $Re=5000$, the other cases lying in between. Thus, in the cases for which the particle stress provides the largest contribution to the momentum transfer (inertial shear-thickening), the diffusion coefficients are well below those of a turbulent flow. In the core region, the particle dispersion is low due to the significant particle accumulation, of the order of that in a laminar flow. However, the turbulent stress is still active (see fig \ref{fig:stress-profile} a) in the intermediate region between the wall and the centerline, which locally increases mixing. This explains why the overall dispersion in the inertial shear thickening regimes is between that of the laminar and turbulent regimes, as shown in figure \ref{fig:Dispersion}. The Reynolds stress and particle stress dominated regimes exhibit therefore different values of the particle dispersion even if they assume the same value of the Bagnold number.

 %

The wall-normal mean square particle displacement is depicted in figure \ref{fig:Dispersion}(b). 
First, we note that  the wall-normal dispersion is lower than its spanwise counterpart due to the presence of the walls. 
The data in figure \ref{fig:Dispersion}(b) asymptotically approach the value $18^2$, the maximum possible wall-normal displacement {when normalized with the particle radius}. As observed for the spanwise dispersion,  the highest and lowest diffusion pertain to the turbulent  and laminar flows, whereas inertial shear-thickening flows display intermediate values. These data on single-particle dispersion  provide additional evidence that two distinct dynamics are at work in the turbulent or inertial shear-thickening regime, despite they can be both classified as inertial Bagnoldian flows.

The values of the spanwise dispersion coefficients {extrapolated for larger $\Delta t$} from the data in figure~\ref{fig:Dispersion} are reported in table \ref{table:Dispersion}. The largest dispersion coefficient is obtained for the case $Re=5000$ and $\Phi=0.05$ where the turbulence activity is the strongest and the lowest one for $Re=500$ and $\Phi=0.05$ in the laminar regime. As mentioned above, the wall-normal mean square displacement is limited by the walls and therefore its slope
does not reach a constant value, corresponding to a well-defined value of the wall-normal dispersion coefficient. 
\\
\begin{table}
\centering
    \begin{tabular}{ | c  | c c c | c c c | c c c | }
    \hline
    Re      & 500 & 2500 & 5000 & 500 & 2500 & 5000 & 500 & 2500 & 5000 \\ \hline
    $\Phi$ & 0.05 & 0.05 & 0.05 & 0.1  & 0.1     & 0.1    & 0.3   & 0.3   & 0.3    \\ \hline
    $\frac{D_x}{\dot{\gamma} a^2}$ & 0.004 & 0.095 & 0.1  &  0.005 &  0.075 & 0.085 & 0.005 & 0.018 & 0.023 \\ \hline
    \end{tabular}
    \caption{\label{table:Dispersion} Dispersion coefficients computed by particle displacements in the spanwise direction.}
\end{table}

 \subsection{Particle-pair statistics}
 We shall also consider the variations of quantities pertaining pairs of particles as a function of the distance between their centers, $r$. 
 As $r$ approaches the particle diameter, the near field interactions become important and collisions occur when $r$ become less than one particle diameter. 
For the details of pair-particle statistics the reader is referred to the appendix of \cite{Collins97} among others. 
Here, we shortly introduce  
the Radial Distribution Function, $g(r)$. In a reference frame with origin at the centre of one particle, the RDF is the averaged number of particle centers  located  in a shell of radius $r$ and $r+dr$ divided by the expected number of particles of a uniform distribution \cite[see][]{Collins00,Gualtieri09}. Formally, $g(r)$ is defined as    
\begin{align}
g(r) = \frac{1}{4 \pi}\frac{d N_r}{d r} \frac{1}{r^2 n_0},
\label{eq:total_stress}
\end{align}
where $N_r$ is the number of particle pairs in a sphere of radius $r$ and $n_0=0.5*N_p(N_p-1)/V$ the density of particle pairs in the volume $V$, with $N_p$ the total number of particles. For small values of $r$, $g(r)$ reveals the intensity of the particle clustering whereas $g(r) \to 1$ when $r \to \infty$ (uniform distribution). 

The dynamics of the particle pair cannot be determined only by the pair distribution function.  Following the study of Sundaram and Collins \cite{Collins97}, we compute the normal relative velocity of the particle pairs as function of $r$. 
Considering particle $i$ and $j$, the normal relative velocity of the particle pair is obtained as the inner product of their relative velocity and relative distance \cite[see][]{Gualtieri12}    
\begin{align}
dv_n(r_{ij}) = (\textbf{u}_i - \textbf{u}_j) \cdot  \frac{(\textbf{r}_i - \textbf{r}_j)}{| (\textbf{r}_i - \textbf{r}_j) |} = (\textbf{u}_i - \textbf{u}_j) \cdot \frac{\textbf{r}_{ij}}{| \textbf{r}_{ij} |}.
\label{eq:total_stress}
\end{align}
The normal relative velocity is a scalar quantity and can be either negative,  $dv^-_n(r_{ab})= dv_n(r_{ab}) \big{|}_{ <0}$, for approaching particles Êor positive, $dv^+_n(r_{ab})=dv_n(r_{ab}) \big{|}_{ > 0}$,  when the two particles depart from each other. The averaged normal relative velocity can be therefore decomposed into  $<dv_n(r)>=<dv^+_n(r)>+<dv^-_n(r)>$. Finally, the collision kernel is obtained as the product of  $g(r)$ and $<dv^-_n(r)>$ \cite[][]{Collins97}, 
\begin{align}
\kappa(r) = g(r)  \cdot <dv^-_n(r)>.
\label{eq:total_stress}
\end{align}

\begin{figure}
\centering{
\includegraphics[width=0.95\linewidth]{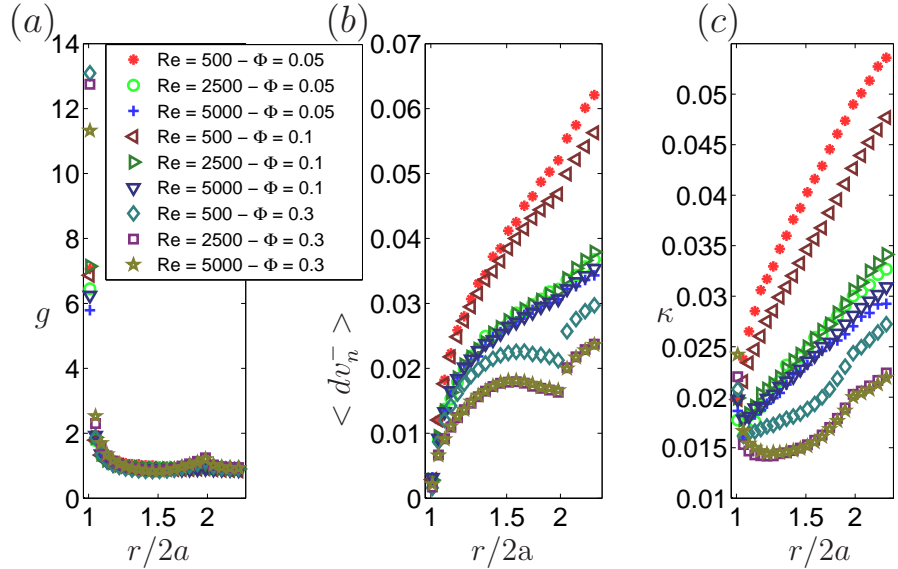}
\put(-45,0){{\large $r/2a$}}
\put(-165,0){{\large $r$/2a}}
\put(-295,0){{\large $r/2a$}}
\put(-330,90){{\large $g$}}
\put(-340,200){{\Large $(a)$}}
\put(-220,50){{\large \begin{turn}{90}$<dv_n^->$\end{turn}}}
\put(-220,200){{\Large $(b)$}}
\put(-95,90){{\large $\kappa$}}
\put(-80,200){{\Large $(c)$}}
\caption{\label{fig:pair} 
(a) Pair distribution function, $g(r)$; (b) Relative normal velocity, $<dv^-_n>$; (c) Collision kernel 
$\kappa(r) = g(r)  \cdot <dv^-_n(r)>$ as a function of the distance between the particle-pair, $r$, for different values of  $Re$ and $\Phi$. }}
\end{figure}

First we compute the Radial Distribution Function, $g(r)$,  the normalized probability of finding pair of particles at distance $r$. The results are shown in figure \ref{fig:pair}(a) for several cases covering the three different regimes. 
For all the cases, the maxima of $g(r)$ occurs at {$r/2a=1$} 
where the particles are in contact and the collision force is active. Note that to compute the RDF, $r$ is discretized in the range of {$r/2a=[0.9, 4]$} and the data for {$r/2a< 1$} are not displayed {(We may have some events where $0.9 < r/2a < 1$ )}. 
Increasing the distance between the particles, $g(r) \to 1$, as expected. The inertial shear-thickening cases are characterized by the highest value of {$g(r/2a=1)$} and by an additional peak {around} {$r/2a=2$} indicating the probability of finding a second layer of the particles. In all the cases, the peak of $g(r)$ decreases slightly when increasing the Reynolds number as the inertia tends to decorrelate the particle paths. 

We then study the normal relative velocity as function of $r$, $<dv_n(r)>$. 
We show the statistics of the negative relative velocity, $<dv^-_n>$, in figure \ref{fig:pair}(b); this observable indicates  the tendency of the particle pairs to approach each other. 
The relative velocity increases {almost}  monotonically with $r$ as the pairs are more likely to approach with higher speed when farther away.  
The highest values of $<dv^-_n>$ occur for the laminar cases and the lowest in the inertial shear-thickening regime, also the densest cases
 (again the relative negative velocity increases smoothly with the Reynolds number for $\Phi=0.3$). The turbulent and inertial shear thickening flows (the Bagnoldian inertial flows) both exhibit values of  $<dv^-_n>$ lower  than the laminar flows. 

The physical mechanism for the reduction of $<dv^-_n>$ is however different. While the turbulent flow tend to homogenize the suspension, the shear induced migration observed in the inertial shear-thickening regime  produces a significant accumulation in the core region where 
 the particles are transported downstream by the core flow at almost constant velocity. 
This suggests that inertia and turbulent eddies determine the local and bulk behaviour of the suspensions at low particle volume fraction, whereas  particle interactions and shear-induced migration governs the behaviour of the flow at high particle volume fractions. The latter effect being almost independent of the flow inertia as discussed above. 

Using single particle statistics, we have shown that the particle dispersion is highest, moderate  and low for the turbulent, inertial shear-thickening and laminar flows respectively. On the other hand, the analyses of the particle-pair statistics, i.e.\ relative velocity, shows the opposite ordering (laminar, turbulent and inertial shear-thickening regimes from high to low). This indicates that these two different aspects of the particle dynamics reflect the presence of the three different bulk regimes in a different fashion.

The collision kernel, the product of $g(r)$ and $<dv^-_n>$, is depicted  in figure \ref{fig:pair}(c). Similarly to the normal relative velocities, this kernel  
increases monotonically with the distance $r$. Interestingly, the kernel assumes similar values when the particles are in contact {($r/2a=1$)} for all the 
cases studied. 

\begin{figure}
\centering{
\includegraphics[width=1.0\linewidth]{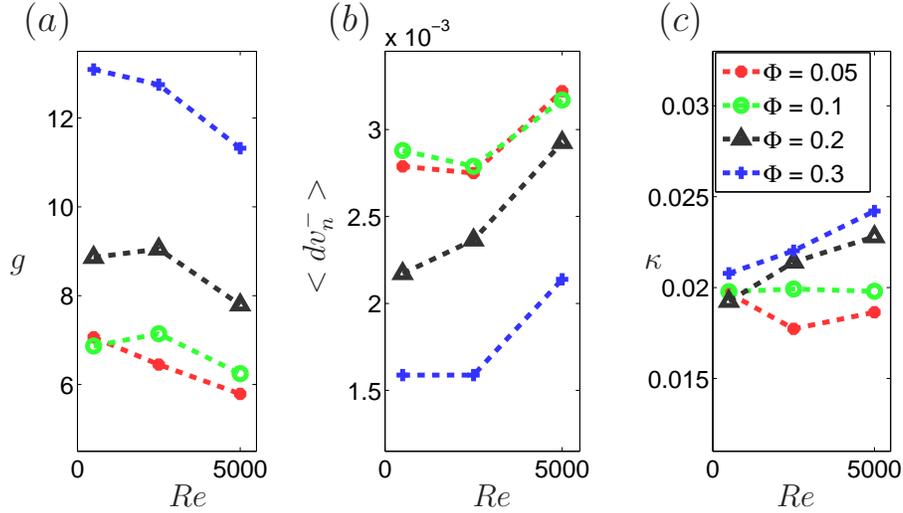}
\put(-55,0){{\large $Re$}}
\put(-170,0){{\large $Re$}}
\put(-285,0){{\large $Re$}}
\put(-345,90){{\large $g$}}
\put(-340,180){{\Large $(a)$}}
\put(-235,80){{\large \begin{turn}{90}$<dv_n^->$\end{turn}}}
\put(-225,180){{\Large $(b)$}}
\put(-105,90){{\large $\kappa$}}
\put(-100,180){{\Large $(c)$}}
\caption{\label{fig:kernel} 
(a) Pair distribution function, $g(r=2a)$; (b) Relative normal velocity, $<dv^-_n> (r=2a)$; (c) Collision kernel 
$\kappa(r=2a) = g(r=2a)  \cdot <dv^-_n(r=2a)>$ versus the Reynolds number for the indicated values of the volume fraction $\Phi$.}}
\end{figure}

To explain the observation that the kernel function has similar values at contact, 
we display in figure \ref{fig:kernel} the separate contribution of $g(r)$ and $<dv_n^-(r)>$ to the kernel at {$r/2a=1$}, where 
the vertical axis are set to cover $\pm50\%$ of the mean values of the data  in each plot to ease a visual comparison. 
The values of {$g(r/2a=1)$ and $<dv_n^-(r/2a=1)>$} are almost independent of the Reynolds number; they do  however increase and decrease  with the particle volume fraction in such a way that their product is almost constant and equal to $0.022 \pm10\%$ for the collision kernel at {$r/2a=1$}. This unique value seems to be valid for all the Reynolds numbers and particle volume fractions studied in the present work as well as those in \cite{Lashgari14}.  Note that the error bar of these statistics is about $1\%$. 
%

As shown above, the dynamics of the particles changes considerably across the channel. To this end, we divide the channel into two regions: region $I$, close to the walls {($0.05 <  z/2h < 0.35$} and $0.65 <  z/2h < 0.95$) and region $II$  the middle of the channel ($0.35 <  z/2h < 0.65$).
Note that the particle centers move in the range $0.05 <  z/2h < 0.95$, as the particle radius $a=0.05$.

%
 %
 %



\begin{figure}
\centering{
\includegraphics[width=0.95\linewidth]{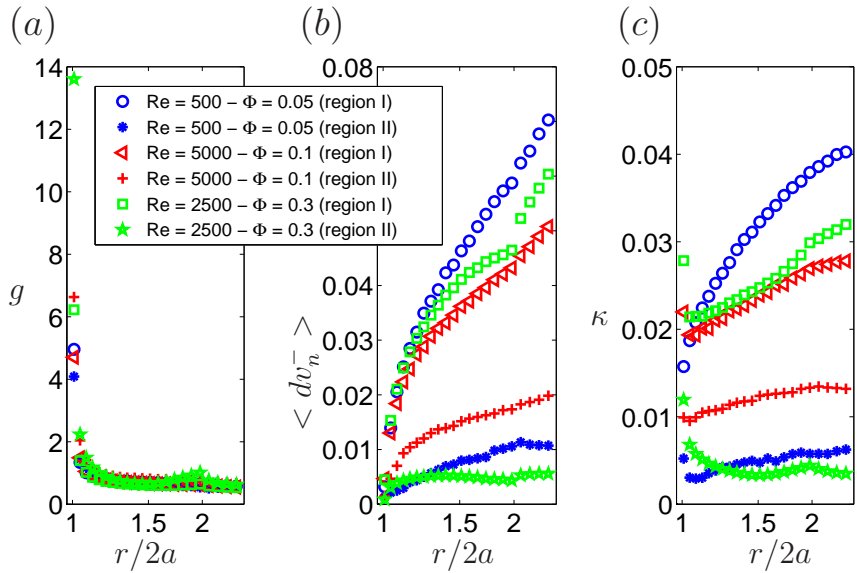}
\put(-50,0){{\large $r/2a$}}
\put(-170,0){{\large $r/2a$}}
\put(-290,0){{\large $r/2a$}}
\put(-330,100){{\large $g$}}
\put(-330,200){{\Large $(a)$}}
\put(-225,50){{\large \begin{turn}{90}$<dv_n^->$\end{turn}}}
\put(-220,200){{\Large $(b)$}}
\put(-110,90){{\large $\kappa$}}
\put(-100,200){{\Large $(c)$}}
\caption{\label{fig:pair_wallcore} 
(a) Pair distribution function (b) Relative normal velocity (a) Collision kernel as a function of distance between the particle-pair, $r$, in the near-wall region $I$ and channel core region $II$ for three configurations. Laminar flow: $Re=500$ $\&$ $\Phi=0.05$, Inertial shear-thickening flow: $Re=2500$ $\&$ $\Phi=0.3$ and turbulent flow: $Re=5000$ $\&$ $\Phi=0.1$ }}
\end{figure}

We thus examine the radial distribution function, negative normal relative velocity and the kernel operator in these regions, limiting the analysis to three flow cases representing the laminar, turbulent and inertial-shear thickening regimes, see figure \ref{fig:pair_wallcore}.
Note that the $g(r)$ is normalized by the total number of particle pairs in each of the two regions. 
The normal relative velocity, see figure \ref{fig:pair_wallcore}(b), is higher in the near-wall region for all the three cases, a fact attributed to the strong background shear, however the difference decreases when the turbulent activity increases.
The collision kernels, see \ref{fig:pair_wallcore}(c), reveal that the kernel of the inertial shear-thickening flow, $Re=2500$ and $\Phi=0.3$, is similar to that of the turbulent flow, $Re=5000$ and $\Phi=0.1$, near the walls and to that of the laminar flow, $Re=500$ and $\Phi=0.05$, in the flow bulk. This is inline with the results in figure \ref{fig:par_rms} pertaining the particle velocity fluctuations.

For the inertial shear-thickening flow, the radial distribution function at contact, {$g(r/2a=1)$}, is higher in region $II$. The difference between region $I$ and $II$ is reduced for the turbulent flow, reflecting  the more uniform particle distribution. The opposite behaviour is observed in the laminar regime where {$g(r/2a=1)$} is instead slightly higher in region $I$. 
\section{Conclusion and remarks} 

We study the flow of suspensions of the finite-size neutrally buoyant particles in a channel, aiming to
connect the local particle  behaviour  to the bulk flow properties. 
The analysis is based on data from direct numerical simulations covering a  wide range of  Reynolds number, $500\le Re \le 5000$, and particle volume fraction, $0 \le \Phi \le 0.3$,  where the particles are rigid spheres with fixed ratio between the particle diameter and channel height of $1/10$.  

The analyses of the stress budget reveals the existence of the three different flow regimes: laminar, turbulent and inertial shear-thickening  depending on which of the stress terms, viscous, Reynolds or particle stress, is the major responsible for the momentum transfer across the channel. 
We show that 
both Reynolds and particle stress dominated flows fall into the Bagnoldian inertial regime \cite{Bagnold54}:
the suspension effective viscosity, i.e.\ the normalised wall shear stress, from the different simulations collapses when
plotted versus the Bagnold number. 
Therefore, turbulent and inertial shear-thickening flows may share the same Bagnold number while the underlying momentum transport and dissipation are distinct.

Examining the particle distribution we show that in the  viscosity dominated laminar flows, characterized by low particle volume fraction and Reynolds number ($\Phi<0.1$ and $Re<1000$), the particles tend to accumulate at certain wall-normal equilibrium positions,  a clear signature of the Segre-Silberberg effect. 
The turbulent particle-laden flow, $\Phi<0.1$ and $Re>1500$, is instead characterised by a more uniform particle distribution due to the mixing by the turbulent eddies. 
At high volume fractions, $\Phi>0.2$, we report a significant migration of the particles toward the channel centerline  for all the Reynolds numbers under investigation, which explains the large contribution of the particle stress in the so-called  inertial shear-thickening regime. 
The particle accumulation in the core region is not necessarily an inertial affect as we observe a negligible variation of the local particle volume fractions when increasing the Reynolds number.   

The mean particle velocity profile becomes more blunt as the flow regime changes from laminar to either turbulent or particle-dominated shear-thickening. 
Interestingly, the 
 velocity fluctuation amplitudes pertaining the inertial shear-thickening are closer to those of the turbulent flow in the near wall region, while they almost overlap to those of the laminar flow in the vicinity of the channel centreline. 
 The particle dynamics  in the inertial shear-thickening regime  appear therefore to share similarities with both the laminar and turbulent flow depending on the wall-normal position. 
 This is further confirmed by examining the spanwise and wall-normal particle dispersion. For the inertial shear-thickening flows the turbulent activity is limited to the near-wall region while in the centre of the channel the particles form a dense layer and are transported by the smooth flow of the carrier fluid. 
 As a result of this,
 the dispersion coefficients of the inertial shear-thickening regime lie in between those of the two other regimes.
 Finally, we note that both the mean and fluctuating particle velocities exhibits slip velocities at the wall.

We further consider the pair particle statistics in the three different regimes. 
In particular, we examine the pair distribution function, $g(r)$, the approaching relative velocities, $<dv_n^->(r)$, and the collision kernel, $\kappa(r)= g(r) <dv_n^->$, as a function of the distance between the particle pairs. The laminar cases show the highest values of $<dv_n^->(r)$ and $\kappa(r)$ while the turbulent flows assume lower values due to the homogeneity created by the turbulent eddies. 
The lowest  values of  $<dv_n^->(r)$ and  $\kappa(r)$ 
are found in the inertial shear-thickening regime, as a consequence 
of the particle packing in the core region. 
Separating the analysis in near-wall and centerline region, we
observe larger values of $<dv_n^->(r)$ and $\kappa(r)$ in the wall region due to the strongest background shear.       

We have therefore demonstrated that the local particle dynamics clearly reflects the existence of the different flow regimes. 
The Bagnold number is shown to correctly predict the bulk flow behavior for the parameter range of this study; however, the details of the momentum transport and dissipation are different in the turbulent and particle-dominated regimes at low and high volume fraction, something which should be considered in any modeling effort. 
We believe future work should consider the role of the particle and fluid inertia and 
the particle dynamics in a mixture of particles of different size.

\section*{Acknowledgements}
This work was supported by the European Research Council Grant No. ERC-2013-CoG-616186, TRITOS and by the Swedish Research Council (VR). The authors acknowledge computer time provided by SNIC (Swedish National Infrastructure for Computing) and the support from the COST
Action MP1305: Flowing matter. {We also acknowledge A. ten Cate for providing the experimental data of figure \ref{fig:Experiment-validation}.}

\section*{References}

\bibliography{mybibfile}

\end{document}